\documentclass[12pt]{article}
\usepackage{epsfig}
\usepackage{axodraw2}
\def\vvpaa{(g_Vg_V^e+g_Ag_A^e)}
\def\vvmaa{(g_Vg_V^e-g_Ag_A^e)}
\def\vapav{(g_Vg_A^e+g_Ag_V^e)}
\def\vamav{(g_Vg_A^e-g_Ag_V^e)}

\def\be{\begin{equation}}
\def\ee{\end{equation}}
\def\beq{\begin{equation}}
\def\eeq{\end{equation}}
\def\beqar{\begin{eqnarray}}
\def\eeqar{\end{eqnarray}}
\def\barr{\begin{array}}
\def\earr{\end{array}}

\def\and{\qquad {\rm and } \qquad}

\def\slp{p \hspace{-1ex}/}

\def\slp{p \hspace{-1ex}/}
\def\sls{s \hspace{-1ex}/}

\def\ttbar{$t \overline{t}~$}

\def\SM{(\vec{s}_+-\vec{s}_-)}
\def\HP{(h_-\vec{s}_+ + h_+\vec{s}_-)}

\def\hp{h_+}

\def\rv{\vec{r}}

\def\rvt{\vec{r}^T}

\def\g5{\gamma_5}
\newcommand{\nc}{\newcommand}
\nc{\jc}{\frac{1}{4}}  \nc{\sll}{S_{LL}}     \nc{\slr}{S_{LR}}
\nc{\srl}{S_{RL}}      \nc{\srr}{S_{RR}}     \nc{\vll}{V_{LL}}
\nc{\vlr}{V_{LR}}      \nc{\vrl}{V_{RL}}     \nc{\vrr}{V_{RR}}
\nc{\tll}{T_{LL}}      \nc{\tlrs}{T_{LR}}    \nc{\trl}{T_{RL}}
\nc{\trr}{T_{RR}}      \nc{\slld}{S_{LL}^D}  \nc{\slrd}{S_{LR}^D}
\nc{\srld}{S_{RL}^D}   \nc{\srrd}{S_{RR}^D}  \nc{\vlld}{V_{LL}^D}
\nc{\vlrd}{V_{LR}^D}   \nc{\vrld}{V_{RL}^D}  \nc{\vrrd}{V_{RR}^D}
\nc{\tlld}{T_{LL}^D}   \nc{\tlrd}{T_{LR}^D}  \nc{\trld}{T_{RL}^D}
\nc{\trrd}{T_{RR}^D}   \nc{\aqde}{\alpha_{qde}}
\nc{\alq}{\alpha_{\ell q}}        \nc{\alqp}{\alpha_{\ell q'}}
\nc{\alqt}{\alpha_{\ell q}^{(3)}} \nc{\alqtc}{\alpha_{\ell
q}^{(3)*}} \nc{\alqj}{\alpha_{\ell q}^{(1)}}
\nc{\alqjc}{\alpha_{\ell q}^{(1)*}} \nc{\aeu}{\alpha_{eu}}
\nc{\alu}{\alpha_{\ell u}} \nc{\aqe}{\alpha_{qe}}
\nc{\ber}{\begin{eqnarray*}} \nc{\enr}{\end{eqnarray*}}
\nc{\jmpb}{(1-\beta)/(1+\beta)} \nc{\wspR}{r}      \nc{\varx}{x}
\nc{\bt}{\beta}

\nc{\non}{\nonumber} \nc{\lspace}{\;\;\;\;\;\;\;\;\;\;}
\nc{\llspace}{\lspace \lspace}
\nc{\jnl}{\frac{1}{{\mit\Lambda}^2}} \nc{\jd}{\frac{1}{2}}
\nc{\comment}[1]{}

\newcommand\splus{(\vec s_+ + \vec s_-)}
\newcommand\sminus{(\vec s_+ - \vec s_-)}
\newcommand\hsplus{(h_- \vec s_+ + h_+ \vec s_-)}
\newcommand\hsminus{(h_- \vec s_+ - h_+ \vec s_-)}

\newcommand\ga{g_A^e}
\newcommand\gv{g_V^e}

\newcommand{\Tone}{$F_1^{pqp}$}
\newcommand{\Ttwo}{$F_1^{psp}$}
\newcommand{\Tthree}{$F_1^{pnp}$}
\newcommand{\Tfour}{$F_1^{ptp}$}
\newcommand{\Tfive}{$F_1^{qsp}$}
\newcommand{\Tsix}{$F_1^{qnp}$}
\newcommand{\Tseven}{$F_1^{qtp}$}
\newcommand{\Teight}{$F_1^{pqs}$}
\newcommand{\Tnine}{$F_1^{pqn}$}
\newcommand{\Tonezero}{$F_1^{pqt}$}

\newcommand{\Toneone}{$F_2^{p}$}
\newcommand{\Tonetwo}{$F_2^{q}$}
\newcommand{\Tonethree}{$F_2^{s}$}
\newcommand{\Tonefour}{$F_2^{n}$}
\newcommand{\Tonefive}{$F_2^{t}$}

\newcommand{\Tonesix}{$PF_1^{pqp}$}
\newcommand{\Toneseven}{$PF_1^{psp}$}
\newcommand{\Tthreesix}{$PF_1^{pnp}$}
\newcommand{\Tthreeseven}{$PF_1^{ptp}$}
\newcommand{\Toneeight}{$PF_1^{qsp}$}
\newcommand{\Tthreeeight}{$PF_1^{qnp}$}
\newcommand{\Tthreenine}{$PF_1^{qtp}$}
\newcommand{\Tonenine}{$PF_1^{pqs}$}
\newcommand{\Ttwozero}{$PF_1^{pqn}$}
\newcommand{\Ttwoone}{$PF_1^{pqt}$}
\newcommand{\Ttwotwo}{$PF_2^{p}$}
\newcommand{\Ttwothree}{$PF_2^{q}$}
\newcommand{\Ttwofour}{$PF_2^{s}$}
\newcommand{\Ttwofive}{$PF_2^{n}$}
\newcommand{\Ttwosix}{$PF_2^{t}$}

\begin{document}

\begin{center}{\Large \bf \boldmath 
Inclusive
spin-momentum analysis and new physics 
at a polarized electron-positron collider}
\vskip 1cm
{B. Ananthanarayan$^a$, Saurabh D. Rindani$^{b}$}
\vskip .5cm

{\it $^a$Centre for High Energy Physics,
Indian Institute of Science\\ Bangalore
560 012, India\\~ \\
$^b$Theoretical Physics Division, Physical Research Laboratory\\
Navrangpura, Ahmedabad 380 009,
India}
\end{center}

\begin{quote}
\begin{abstract}
We consider the momentum distribution and the polarization of an 
inclusive heavy fermion in a process assumed to arise from
standard-model (SM) 
$s$-channel exchange of a virtual $\gamma$ or $Z$ 
with a further contribution from physics
beyond the standard model involving $s$-channel exchanges. The
interference of the new physics amplitude with the SM  $\gamma$ or $Z$
exchange amplitude is expressed 
entirely in terms
of the space-time signature of such new physics.  
Transverse as well as longitudinal polarizations of the electron and
positron beams are taken into account. Similarly, we consider the cases
of the polarization of the observed final-state fermion along
longitudinal and two transverse spin-quantization axes which
are required for a full reconstruction of the spin dependence of the
process.
We show how these model-independent distributions can be used to 
deduce some general properties of the
nature of the interaction and
some of their properties in prior
work which made use of spin-momentum correlations. 
\end{abstract}
\end{quote}
\newpage
\section{Introduction}\label{intro}

The proposed International Linear Collider (ILC)~\cite{ILC} which could 
collide $e^+$ and $e^-$
at a centre-of-mass energy of several hundred GeV, if built, would serve
as an instrument for precision measurements of various parameters
underlying particle physics and the dedicated study has published
a five-volume Technical Design Report (for the physics part, see
ref.~\cite{LC_SOU} and for the detector see ref.~\cite{ILC_DET}).
The purpose of the ILC, and indeed of
other proposed high energy $e^+e^-$ colliders,
such as the Compact Linear Collider (CLIC)~\cite{CLIC1,CLIC2}
the Future Circular Collider (FCC-ee)~\cite{FCCee1,FCCee2}
and the Circular Electron Positron Collider (CEPC)~\cite{CEPC}
is to study the
properties of the Standard Model (SM) at high precision in order to
validate its predictions as well as to find deviations, if any, and 
to discover particles and interactions that lie Beyond the
Standard Model (BSM).  Deviations from SM predictions would arise
because of virtual loop effects of particles too heavy to be produced,
or indeed due to new interactions which would give rise to terms
in the low-energy effective action modifying interaction
vertices.  Amplitudes from such vertices could interfere with SM amplitudes
and produce deviations from its predictions, and could possibly give
rise to correlations that are forbidden by the symmetries of the SM
when SM particles are observed in the detectors with high-precision
measurements of their kinematic and other properties.
A dedicated study on the benefits of a strong beam polarization programme,
of either or both beams, as well as the benefits of transverse and
longitudinal beam polarization has also been carried out some years
ago in the context of the ILC~\cite{POWER}.  An important new compendium
of physics at the ILC is the review, ref.~\cite{newILC}.

A useful approach that has been applied in the
context of BSM physics searches at $e^+e^-$ colliders relies on the
classification of new physics in terms of its space-time
transformation properties using, e.g., one-particle~\cite{BASDR1} inclusive distributions 
$e^+ e^-\to h(p)  X$, where $h$ denotes a
particle that is detected, and $p$ is its momentum.
The new physics is lumped into `structure functions' that
are inspired by analysis used in deep-inelastic scattering.
It has also been extended in the context of   
a two-particle inclusive process~\cite{BASDR2} 
$e^+ e^-\to h_1(p_1) h_2(p_2) X$, where $h_1$ and $h_2$ denote two
particles that are detected, and $p_1$ and $p_2$ are their
respective momenta and is denoted as the basic process (I).
The two-particle inclusive process is 
depicted in Fig.~\ref{bfig1}, while
the one-particle process can be considered a special case where 
only one of the particles
is detected and the other is included in $X$.

\begin{center}
\begin{figure}[htb] \label{bfig1}
\begin{picture}(500,100)(-50,0)
\ArrowLine(10,100)(41.5,60.5)
\Text(-10,90)[]{$e^+(p_+)$}
\Text(85,75)[]{$h_1(p_1)$}
\Text(85,40)[]{$h_2(p_2)$}
\Text(85,60)[]{$X$}
\Text(135,60)[]{$=$}
\Text(300,60)[]{$+$}
{\GOval(45,60)(5,5)(0){0.5}}
\ArrowLine(10,20)(39.5,59.5)
\Text(-10,40)[]{$e^-(p_-)$}
\ArrowLine(49,61)(80,100)
\ArrowLine(49,59)(80,20)
\ArrowLine(50,60)(80,60)
\ArrowLine(160,100)(190,60)
\ArrowLine(160,20)(190,60)
\Photon(190,60)(250,60){2}{6}
\Text(220,80)[]{$\gamma, \, Z$}
\ArrowLine(250,60)(280,100)
\ArrowLine(250,60)(280,20)
\ArrowLine(250,60)(280,60)
\ArrowLine(310,100)(340,60)
\ArrowLine(310,20)(340,60)
\ArrowLine(342,60)(372,100)
\ArrowLine(342,60)(372,20)
\ArrowLine(342,60)(372,60)
{\GOval(341,60)(2,2)(0){0}}
\Text(343,45)[]{\small NP}
\end{picture}
\caption{The basic process (I)}
\end{figure}
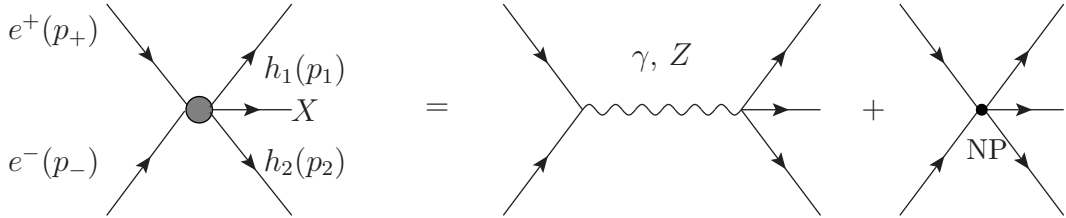
\end{center}

This approach is model independent, and is based only on
Lorentz covariance for deriving the most general form of
one-particle and two-particle kinematic distributions.
It was found that the two-particle case provides more information
than the single-particle case as discussed in detail in 
ref.~\cite{BASDR2}, and 
in principle, this could be extended to an
$n$-particle inclusive framework, with a rapid rise in complexity.
Our formalism is restricted to envisaging new physics only through an
$s$-channel exchange, i.e., that
(i)  the SM contribution is  assumed
to be through the tree-level exchange of a virtual photon and a virtual
$Z$, and that
(ii) the BSM effects could arise through the
exchange of a new particle like a new gauge boson $Z'$, or through the
exchange of $Z$, but a with a BSM vertex or a SM loop producing the 
final state in question, or through a new scalar or tensor exchange in
the $s$ channel.  
Our work above is an extension of the  work of Dass and Ross~\cite{DR1,DR2} 
that had been performed in the context of $\gamma$ contributing
to the $s$-channel production, probing the then undiscovered neutral
current.  As discussed extensively in refs.~\cite{BASDR1,BASDR2},
our work in practice is the inclusion of 
$Z$ in the s-channel, in addition to $\gamma$, and 
where now it is BSM physics that we intend to fingerprint.
Moreover, the results can be applied to a more general situation where
the interference need not be between SM and BSM amplitudes, but any two
amplitudes, one of which is characterized by the exchange of a spin-1
particle, and the other characterized by scalar, pseudoscalar, vector,
axial-vector or tensor interactions.

Many studies of such manifestations of BSM physics rely on the measurement
of an exclusive final state for which there are definite predictions
in the SM, and/or definite predictions within the framework of
effective Lagrangians, or effective BSM vertices. An early work in
this regard in the context of the ILC is ref.~\cite{Ananthanarayan:2003wi},
where it was shown that transverse polarization plays a key role in
uncovering CP violation due to BSM physics due to scalar (S),
pseudo-scalar (P) and tensor (T) type interactions, when no
spins are measured.  This work was inspired partly by even earlier
work done for LEP energies, see ref.~\cite{BurgessRobinson}.

The question then arises as to how one may be able to probe
BSM physics further with one-particle inclusive distributions,
in the event that its spin has been measured.  Keeping in mind
that spin measurement is actually performed by further decays
of the particle in question, such a scenario is really a
quasi-one-particle inclusive process.  Nevertheless,
the availability of a second final-state momentum vector
as in the two-particle inclusive case is what renders it a more
powerful probe.  On the other hand, a single-particle
inclusive measurement with the measurement of spin of
the particle along a specific quantization axis may provide
a second vector and thereby play an important role in uncovering
BSM physics.  Whereas in ref.~\cite{DR2}, the possibility of measuring the
spin of the particle in a one-particle inclusive measurement
$e^+ e^-\to h(p,s)  X$, where $h$ and $s$ denote the  SM
particle that are detected, and $h$ and $s$ are its
momentum and spin respectively,
has been considered, it had not been 
considered in ref.~\cite{BASDR2}.  This is denoted as the basic process (II)
and is depicted in Fig. 2.

\begin{center}
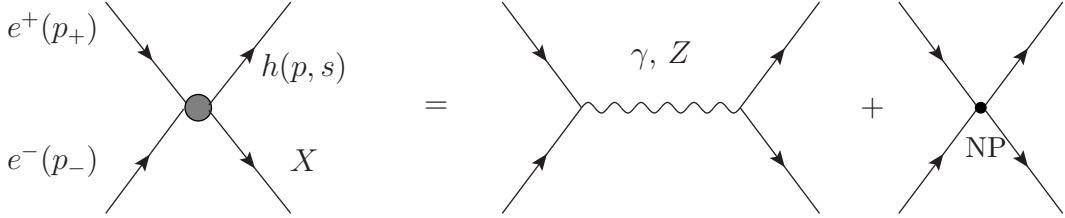
\begin{figure}[htb] 
\begin{picture}(500,100)(-50,0)
\ArrowLine(10,100)(41.5,60.5)
\Text(-10,90)[]{$e^+(p_+)$}
\Text(85,75)[]{$h(p,s)$}
\Text(85,40)[]{$X$}
\Text(135,60)[]{$=$}
\Text(300,60)[]{$+$}
{\GOval(45,60)(5,5)(0){0.5}}
\ArrowLine(10,20)(39.5,59.5)
\Text(-10,40)[]{$e^-(p_-)$}
\ArrowLine(49,61)(80,100)
\ArrowLine(49,59)(80,20)
\ArrowLine(160,100)(190,60)
\ArrowLine(160,20)(190,60)
\Photon(190,60)(250,60){2}{6}
\Text(220,80)[]{$\gamma, \, Z$}
\ArrowLine(250,60)(280,100)
\ArrowLine(250,60)(280,20)
\ArrowLine(310,100)(340,60)
\ArrowLine(310,20)(340,60)
\ArrowLine(342,60)(372,100)
\ArrowLine(342,60)(372,20)
{\GOval(341,60)(2,2)(0){0}}
\Text(343,45)[]{\small NP}
\end{picture}
\caption{The basic process (II)}
\end{figure}\label{basicfig2}
\end{center}

We also note here that in
the recent past, numerous investigations have been made in the context of exclusive processes at the ILC, where it has been shown that the measurement of
the final-state spin can also be an excellent probe of BSM physics.  We had considered specific exclusive processes and had concluded that many
types of BSM interactions reveal themselves only when the spin of
one of the final state particles is resolved.   For instance,
in order to separately resolve BSM contributions from scalar and
tensor type couplings in $e^+e^-$ collisions with transversely
polarized beams, one has to resolve the spin of the top-quark
in \ttbar production~\cite{Ananthanarayan:2010xs,Ananthanarayan:2012ir}.\footnote{Processes not covered here are those were the
SM production goes through t- and u- channel diagrams, as in the
case of vector boson production.  Nevertheless, it may be pointed out that even in this context certain anomalous triple-gauge boson
couplings in $\gamma Z$ production also become visible only with
the resolution of the spin of the final state bosons, see 
refs.~\cite{Ananthanarayan:2011fr,Ananthanarayan:2014sea,Ritesh}.}
Early work on the necessity to resolve the final state spins in
the context of $\tau^+\tau^-$ production at significantly lower energies
to probe the presence of anomalously large magnetic moments and
possible electric dipole moments of the $\tau$-lepton are refs.~\cite{Couture1,Couture2}.
(For a general and interesting discussion see ref.~\cite{HoogeveenStodolsky}.)
In the present work, for purposes of illustration, we introduce these
sources of BSM physics to provide a concrete framework wherein we
can make some remarks about the resulting structure functions derivable from
such an exclusive process.

It is usual practice to study the dependence of a process on the spin of a produced
particle by restricting to a single spin quantization axis, typically,
the momentum direction of the particle. In this case, what is accessible
is the probability of production of the particle with a definite
helicity. However, this corresponds to only the diagonal element of the
spin density matrix. In order to study the full spin structure of
amplitudes, one needs also off-diagonal elements of the spin density
matrix, or equivalently, the polarization information for two other
mutually orthogonal spin quantization axes. This approach has been
advocated earlier, for example, in refs.~\cite{Harlander,Ritesh2} in the
context of top-pair production at an $e^+e^-$ collider and in~\cite{AS} 
for single-top production at the Large Hadron Collider.  Single-top production
itself is an interesting process in itself; for a review, see ref.~\cite{BoosDudko}.
Single-top production at CLIC has been studied in  ref.~\cite{Koksal}.

Keeping in mind these considerations,
for the purposes of this work  we confine
ourselves to an inclusive, massive spin-1/2 fermion,
where 
we now employ the three suitably chosen axes explicitly.  
We note here that the considerations of ref.~\cite{DR2} remained
general in the choice of the spin quantization axis.  In
the present work, we present results for the three different
quantization axes. In practice,
this is made possible by the fact that 
the two types of processes 
are closely related:  the single-particle inclusive process with
spin measurement is closely related to the two-particle inclusive
process with suitable identification of vectors entering the definition
of the structure functions.  Thus, by employing the standard techniques
as in ref.~\cite{BASDR1,BASDR2,DR1,DR2} we can proceed with the analysis of the
single-particle inclusive measurement with spin resolution.
As in our earlier work, 
significant new features arise
due to the presence of the axial-vector coupling of the $Z$
to the electron, a feature missing in a vector theory like QED,
as in the considerations of ref.~\cite{DR2}, and an extensive
discussion can be provided on the features of the correlations
for the three specific quantization axes.
In all considerations of the top-quark spin resolution
at the LHC as ref.~\cite{AS} or at the ILC, as well as in
$\tau-$spin reconstruction as in the work discussed
here, it can only be done from the distributions of its decay products
and typically taken in the rest frame of the top-quark, by looking
at the angular distribution of a decay product about the quantization
axis.  This, of course, is independent of the
environment in which the top-quark is produced, whether it is
in a hadron collider or an $e^+e^-$ collider, and whether or
not in the hadron collider it is pair or singly produced.
Analogous considerations apply also to the $\tau$-lepton.
For reviews on approaches to these issues, see
refs.~\cite{Nelson,Kuhn,Kitano,Khiem,Li}.

As in the past, once a general discussion is provided
for an inclusive final state, it may be readily applied to
exclusive final states as well, thereby providing a framework
for discussing several processes of interest.
The expectations from our general model-independent analysis is compared for 
some specific processes with the results obtained
earlier for those processes. Our approach would thus be useful to derive
general results for newer processes which fall within the framework
described above. Thus, what is presented here is the result of a 
detailed calculation for each individual process.

We also note that many of the considerations that have been
spelt out for the ILC also apply to the other planned facilities,
namely CLIC, FCee and CEPC.

The structure of this paper is as follows:
In the next section we include some preliminaries about the inclusive
process, the kinematics and a discussion on the choice of spin
quantization axes.
In Sec.~\ref{correlations} we present a computation of the spin-momentum
correlations resulting from the presence of structure functions
that characterize the new physics.  Our results here are presented
in the form of results arising from the computation of a trace that
encodes the leptonic tensor as well as the new physics encoded in
a tensor constructed out of the momenta of the observed final-state
particles (what is known as a `hadronic' tensor, for historical reasons,
since the term arose at a time 
when the final state consisted largely of hadrons).  
These tables provide the
analogue for the SM and new physics, of what was 
provided by Dass and Ross~\cite{DR2}
for QED and neutral currents.  In Sec.~\ref{characterization} 
we discuss the CP and T properties of correlations for different classes
of inclusive and exclusive final states. 
We provide a discussion on the the polarization dependence of the
correlations in different cases.
In Sec.~\ref{exclusive} we will specialize to specific examples of 
processes, into which our approach 
 can give significant insight.  
In Sec.~\ref{conclusions} we present our conclusions and discuss
prospects for extension of the present framework to account for classes
of BSM interactions not presently covered.

\section{The process and kinematics}

We consider the two-particle inclusive process and the one-particle
spin-resolved process
\begin{equation}\label{process}
e^-(p_-) + e^+(p_+) \to h(p,s) X,          
\end{equation}
where $h$ is the final-state particles whose momentum $p$ and 
spin $s$ are measured,
$X$ is an  inclusive state.
The process is assumed to occur through an $s$-channel
exchange of a photon and a $Z$ in the SM, and through a  new current whose
coupling to $e^+e^-$ can be of the type $V,A$, or $S,P$, or $T$.
Since we will deal with a general case without specifying the nature or
couplings of $h$, 
we do not attempt to write the amplitude for the process
(\ref{process}). We will only obtain the general form, for each case of
the new coupling, of the contribution to the angular distribution of $h$ 
from the interference of the SM amplitude with the new physics
amplitude. It might be clarified here that even though we use the term
``inclusive'' implying that no measurement is made on the state $X$, in
practice it may be that the state $X$ is restricted to a concrete one-particle
or two-particle state which is detected. In such a case the sum is not
over all possible states $X$. Nevertheless, the momenta of the few
particles in the state $X$ are assumed to be integrated over, so that
there is a gain in statistics as compared to a completely exclusive
measurement. The angular distributions we calculate hold also for such a
case, except that structure functions would depend on the states
included in $X$.

The following symbols have been used by us in various stages of
the computations and we present here a comprehensive list of these 
definitions.  We define: $q=p_-+p_+$, $\vec K\equiv (\vec{p}_- - \vec{p}_+)/2= E \hat{z}$,
where $\hat{z}$ is a unit vector
in the z-direction, $E$ is the beam energy,
and $\vec{s}_\pm$ lie in the x-y plane.

We now turn to the important question of the choice of
three linearly independent vectors which will define the
quantization axes.  Although the decay distributions of
the top-quark are correlated to the spin in the top-quark
rest frame, our choice of vectors is in the laboratory, or
$e^+e^-$ c.m. frame.   It is assumed that all the kinematic
information would be available which would allow one to
construct any quantity of interest for the event sample.
In particular, it may be noted that this choice would suffice
for the full analysis of the top-quark polarization for
which the SM would have definite predictions, and could also
be used in other contexts such as anomalous couplings, or
any kind of BSM physics.

In the $e^+e^-$ centre-of-mass
frame, the spin vectors have components given by:
\begin{equation}\label{spinvectors}
 s^\mu \equiv \frac{1}{m} ( | \vec p |, E_p \hat p ),
\end{equation}
\begin{equation}\label{spinvectorn}
 n^\mu \equiv \frac{1}{2E\sin\theta}( 0, \hat p \times (\vec p_- - \vec
p_+) ),
\end{equation}
\begin{equation}\label{spinvectort}
 t^\mu \equiv ( 0, - \hat p \times \vec n ),
\end{equation}
where $P = \vert \vec p \vert $ and $E_p = \sqrt{P^2 + m^2}$.
In covariant notation,
\begin{equation}
 n_\mu \equiv \frac{1}{\sqrt{q^2(4 (p\cdot p_-)( p\cdot p_+) - m^2 q^2)}}\epsilon_{\mu\nu\alpha\beta} p^\nu 
(p_-+p_+)^\alpha (p_+-p_-)^\beta , 
\end{equation}
\begin{equation}
\begin{array}{ll}\displaystyle
t_\mu \equiv\frac{1}{\sqrt{q^2(4 (p\cdot p_-)( p\cdot p_+) - m^2 q^2)}}& \left[
\displaystyle \frac{ p\cdot
(p_- - p_+)q^2}{\sqrt{(p\cdot q)^2 - m^2 q^2}} p_\mu +
\sqrt{(p\cdot q)^2 - m^2q^2} \, (p_- - p_+)_\mu \right. \\&\left.  + 
\displaystyle
\frac{p\cdot q\,p\cdot (p_+ - p_-)}{\sqrt{(p\cdot q)^2 - m^2 q^2}}
(p_- + p_+)_\mu \right].
\end{array}
\end{equation}
Note that $\theta$ is the angle of $h$ with respect to the beam-direction,
and $\phi$ is the azimuthal angle where the $x$-axis is chosen as
the direction of the transverse polarization of the $e^+$ and $e^-$ beams.  Note that the symbol $h(p,s)$ is a generic symbol,
where the spin $s$ could stand for the measurement of the
spin along any one of the three quantization axes of choice.

The three-vectors $\vec n$ and $\vec t$ in the laboratory frame 
are actually quite simple:
\begin{equation}\vec{n} \equiv (\sin\phi, - \cos\phi, 0)
\end{equation}
 and
\begin{equation}
\vec{t} \equiv ( - \cos\theta  \cos\phi, -\cos\theta \sin\phi,
\sin\theta).
\end{equation}
$\vec n$ is along a direction perpendicular to both the momentum $\vec
p$ of $h$ and the beam direction, see for example ref.~\cite{Haber:1994pe}. 
On the other hand, $\vec t$ is in the
plane of the beam direction and $\vec p$, though perpendicular to the
latter.
For ease of visualization, we have represented the vectors in
Fig. 3.

{\centering
\begin{center}
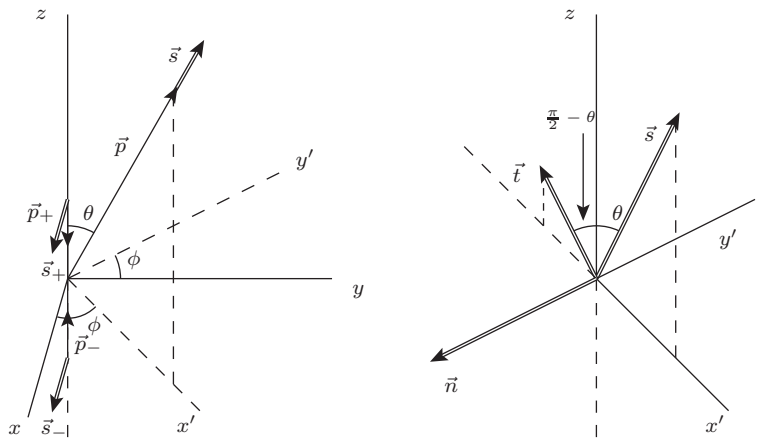
\begin{figure}[htb]
\begin{picture}(-200,200)(-125,0)
\Line(0,0)(0,100)
\DashLine(0,0)(0,-60){5}
\ArrowLine(0,30)(0,0)
\Line[arrow,arrowpos=1,double,sep=1,arrowscale=1.0](40,70)(50,87.5)
\LongArrow(0,0)(40,70)
\Line[arrow,arrowpos=1,double,sep=1,arrowscale=1.0](0,30)(-5,12.5)
\ArrowLine(0,-30)(0,0)
\Line[arrow,arrowpos=1,double,sep=1,arrowscale=1.0](0,-30)(-5,-47.5)
\Line(0,0)(100,0)
\Line(0,0)(-15,-52.5)
\DashLine(0,0)(50,-50){5}
\DashLine(0,0)(80,40){5}
\DashLine(40,70)(40,-40){5}
\Text(-10,25)[]{{\scriptsize$\vec{p}_+$}}
\Text(8,-25)[]{{\scriptsize$\vec{p}_-$}}
\Text(20,50)[]{{\scriptsize$\vec{p}$}}
\Text(8,25)[]{{\scriptsize$\theta$}}
\Text(-5,3)[]{{\scriptsize$\vec{s}_+$}}
\Text(-5,-57)[]{{\scriptsize$\vec{s}_-$}}
\Text(40,85)[]{{\scriptsize$\vec{s}$}}
\Text(10,-18)[]{{\scriptsize$\phi$}}
\CArc(0,0)(15,255,315)
\CArc(0,0)(20,60,90)
\CArc(0,0)(20,0,26)
\Text(25,7)[]{{\scriptsize$\phi$}}
\Text(-20,-57)[]{{\scriptsize$x$}}
\Text(45,-55)[]{{\scriptsize$x'$}}
\Text(-10,100)[]{{\scriptsize$z$}}
\Text(110,-5)[]{{\scriptsize$y$}}
\Text(90,45)[]{{\scriptsize$y'$}}
\Line(200,0)(200,100)
\DashLine(200,0)(200,-60){5}
\Line[arrow,arrowpos=1,double,sep=1,arrowscale=1.0](200,0)(230,60)
\Line(200,0)(250,-50)
\Line[arrow,arrowpos=1,double,sep=1,arrowscale=1.0](200,0)(180,40)
\DashLine(200,0)(150,50){5}
\Line(200,0)(260,30)
\Line[arrow,arrowpos=1,double,sep=1,arrowscale=1.0](200,0)(140,-30)
\DashLine(180,40)(180,20){3}
\DashLine(230,60)(230,-30){5}
\Text(208,25)[]{{\scriptsize$\theta$}}
\Text(220,55)[]{{\scriptsize$\vec{s}$}}
\CArc(200,0)(20,65,90)
\CArc(200,0)(20,90,115)
\Text(145,-40)[]{{\scriptsize$\vec{n}$}}
\Text(170,40)[]{{\scriptsize$\vec{t}$}}
\Text(190,60)[]{{\tiny$\frac{\pi}{2}-\theta$}}
\LongArrow(195,55)(195,25)
\Text(245,-55)[]{{\scriptsize$x'$}}
\Text(190,100)[]{{\scriptsize$z$}}
\Text(250,15)[]{{\scriptsize$y'$}}
\end{picture}
\vskip 3cm
\caption{Representation of the momentum and spin vectors in the laboratory frame.  
The left panel depicts the electron and positron momenta, respectively, $\vec p_-$ and $\vec p_+$, 
their respective transverse spin vectors $\vec s_-$ and $\vec s_+$, 
which lie along the positive and negative $x$-axis respectively, 
the momentum $\vec p$ of the detected particle and the spin-quantization axis 
that is denoted by $\vec s$.  The right panel depicts the three different
spin-quantization axes $\vec s$, $\vec n$ and $\vec t$ defined in the text. In both panels, 
the axes $x'$ and $y'$ denote axes obtained by rotation of the $x$ and $y$ axes about the $z$ axes through an angle $\phi$.}
\end{figure} \label{spinvectorsfig}
\end{center}
}

As in the past, we calculate the relevant factor in
the interference between the standard model currents with the
BSM currents as
\begin{equation}\label{trace}
{\rm Tr}[(1-\g5 \hp + \g5 \sls_+)\slp_+\gamma_\mu(g_V^e-g_A^e \gamma_5)
(1+\g5 h_-+\g5 \sls_-)\slp_-\Gamma_i]H^{i\mu }.
\end{equation}
Here $g_V^e, g_A^e$ are the vector and axial-vector couplings of the
photon or $Z$ to the electron current, and $\Gamma_i$ is the
corresponding coupling to the new-physics current, 
$h_{\pm}$ are the
helicities (in units of $\frac{1}{2}$)
of $e^{\pm}$, and $s_{\pm}$ are respectively their transverse polarizations.
For ease of comparison, we have sought to stay with the notation
of refs.~\cite{DR1,DR2}, with some exceptions which we
spell out when necessary.  We should of course add the
contributions coming from photon exchange and $Z$ exchange, with the
appropriate propagator factors. However, we give here the results for
$Z$ exchange, from which the case of photon exchange
 can be deduced as a special
case. The tensor $H^{i\mu }$ stands for the interference between the
couplings of the final state to the SM current and the new-physics
current, summed over final-state polarizations, and over the phase space
of the unobserved particles $X$. It is only a function of the the
momenta $q$ , $p$ and $s$ (or $n$ or $t$). The implied summation over $i$
corresponds to a sum over the forms $V, A, S, P, T$, together with any
Lorentz indices that these may entail.

\section{Computation of correlations}\label{correlations}

We now determine the forms of the matrices $\Gamma_i$ and the
tensors $H^{i\mu }$ in the various
cases, using only Lorentz covariance properties.
Our additional currents are as in refs.~\cite{DR1,DR2},
except for the sign of $g_A$ in the following.  We explicitly
note that in our convention is $\epsilon^{0123}=+1$.
We set the electron mass to zero.  
Consider now the three cases where the BSM physics could
be of the scalar and pseudoscalar type, vector and axial-vector
type or tensor type. Note that in each case, $H^{i\mu }$ can be
independent of the spin vector ($s$, $n$ or $t$), or linearly dependent
on the spin vector. The linear dependence can arise either from the
spin vector entering the tensor structure or from a simple
multiplicative factor $q\cdot s$ ($q\cdot n$, $q\cdot t$ being zero in the 
centre-of-mass frame). We explicitly include the tensors which involve
the spin vector. But we do not show the spin vector entering through a
factor of $q\cdot s$, as this would have the same distribution as the
spin-independent tensor. It is thus understood that in what follows,
each spin-independent structure function, say $F$, should be actually
replaced by $F + F' (q\cdot s)$, where $F'$ is another structure
function. 

\bigskip

\noindent\underline {1. Scalar and pseudoscalar case}:
\smallskip

In this case, there is
no free Lorentz index for the leptonic coupling.
Consequently, we can write it as
\begin{equation}
\Gamma = g_S + i g_P \gamma_5.
\end{equation}
The tensor $H^{i\mu }$ for this case has only one index, viz., $\mu$.
Hence the most general form 
for $H^S_\mu$ is\footnote{The form for $H^S_\mu=(r_\mu - q_\mu {r\cdot q \over q^2}) F^r$ 
which is the definition adopted in ref.~\cite{DR2},
is also permissible, since when $r=p$, and
since $p - q {p\cdot q \over q^2}$ is a current conserving
combination, and the second term does not contribute.}:
\begin{equation}\label{scalarH}
H^{S}_\mu = r_\mu F^r,
\end{equation}
where
$r$ is chosen from $p$, and the spin vectors $s$, $n$ and $t$,
corresponding respectively to longitudinal polarization, transverse
polarization perpendicular to the production plane, and transverse
polarization in the production plane.  Here $F^r$ denotes the relevant
structure functions we encounter and is a function of invariants
$q^2$ and $p\cdot q$.  In fact, all 
the structure functions introduced in the above,
as well as those to be introduced in the following are functions of the
same Lorentz invariants $q^2$ and $p\cdot q$.
The dependence of the functions on $q^2$ and $p\cdot q$
encodes the dynamics of the BSM interactions. In particular, they would
contain propagators and form factors occurring in the BSM
amplitudes.
It may be noted that these definitions
can result in an unconventional phase for the spin vectors, and will have
implications to our analysis of spin-momentum distributions and their
properties implied by the CPT theorem:  it would imply that relative to
the momentum the spin vectors would have required an additional factor
of $i$ in the definition of the structure functions in the usual correspondence
between CPT$=-1$ distributions and the appearance of imaginary parts of
these structure functions~\cite{Rindani:ad1994}. (For another useful
review, see ref.~\cite{Bernreuther}).

\bigskip

\noindent\underline {2. Vector and axial-vector case}:
\smallskip

The leptonic
coupling for this case can be written as
\begin{equation}
\Gamma_\nu = \gamma_\nu (g_V -  g_A \gamma_5).
\end{equation}
Note that we differ from Dass and Ross \cite{DR1,DR2} in the sign of the
$g_A$ term,   in order to be in line with the convention for the
standard neutral current coupling of the SM, which was established
well after the work in refs.~\cite{DR1,DR2}. 
The tensor $H$ for this case has two indices, and can be written as
\begin{eqnarray}\label{Hvector}
&
H^V_{\mu\nu} =  -g_{\mu\nu} W_1
+ {1\over 2}(r_\mu w_\nu+r_\nu w_\mu) W_2^{rw}
& \nonumber \\
& + \epsilon_{\mu\nu\alpha\beta}u^{\alpha}v^\beta W_3^{uv}
+{1\over 2}(p_{\mu} n_{\nu} - p_{\nu} n_{\mu}) W_4
\end{eqnarray}
where $W_1, W_2^{rw}, W_3^{uv}, W_4$ are invariant functions,
and $r$, $w$ can be chosen from $p$, $s$, $n$
and $t$,
and $u,v$ can be chosen from $p$, $q$, $s$, $n$ and $t$, with the
condition that the tensor be at most linear in the spin vector.
As compared to the one-particle exclusive case,
there is an additional tensor structure with structure function $W_4$,
which requires two vectors, being antisymmetric. The only non-zero
contribution is in the case when two vectors are $p$ and $n$.

\bigskip 

\noindent\underline {3. Tensor case}:
\smallskip

In the tensor case,
the leptonic coupling is
\begin{equation}\label{tensorG}
\Gamma_{\rho\tau} = g_T \sigma_{\rho\tau}.
\end{equation}
The tensor $H$ for this case can be written in terms of the four
invariant functions $F_1$, $F_2$, $PF_1$, $PF_2$ as
\begin{equation}
\begin{array}{lcl}\label{tensorH}
H^T_{\mu\rho\tau}& = & (r_\rho  u_\tau - r_\tau u_\rho ) w_\mu F_1^{ruw}
+ ( g_{\rho\mu} u_\tau - g_{\tau\mu} u_\rho ) F_2^u
\\&& + \epsilon_{\rho\tau\alpha\beta} r^\alpha u^\beta w_\mu
PF_1^{ruw}
+ \epsilon_{\rho\tau\mu\alpha} u^\alpha PF_2^u,
\end{array}
\end{equation}
where $w$ is chosen from $p$, $s$, $n$ and $t$, 
$r$ from $p$ and $q$, and
$u$ from $p$, $q$, $s$, $n$ and $t$. 
These choices of vectors for $r$, $w$ and
$u$ give a complete set of independent tensors. The use of vectors other
than covered by the choices would result in tensors which are
combinations of tensors described by eq. (\ref{tensorH}). Details can be
found in \cite{DR2}.

We next substitute the leptonic vertices $\Gamma$ and the respective
tensors $H^i$ in  (\ref{trace}), and
evaluate the trace in each case. We present the results in Tables
1-4. The structure functions accompanying
the tensors which depend on spin (i.e. contain one of
the vectors $\vec s$, $\vec n$ and $\vec t$), would occur in the the
spin-dependent differential cross section with a factor $\lambda_s$,
$\lambda_n$ or $\lambda_t$, each taking the value $+1$ or $-1$, denoting
the spin projection along the respective spin vector $\vec s$, $\vec n$
or $\vec t$. 

A superscript $T$ on a vector is used to denote its component transverse
with respect to the $e^+e^-$ beam directions. For example, $\rvt = \rv -
\rv\cdot\hat{z}\,\hat{z}$, and similarly for other vectors.
Tables \ref{scalartable}, \ref{vectortable}, \ref{retensortable} and 
\ref{imtensortable}
are respectively for cases of scalar-pseudoscalar,
vector-axial-vector, real and imaginary parts of tensor couplings respectively.

Since our present case is that of a single particle being measured,
there is only one momentum $p$. However, there is one more vector, viz. the
spin vector.   These are the three possible spin quantization axes given
by $s^\mu,\, n^\mu$ and $t^\mu$ and a full evaluation of the resulting
correlations is given in Tables 1-4.

\begin{table}
\begin{center}
\begin{tabular}{|c|c|}
\hline
Structure function & Correlation \\
\hline
   ${\rm Im} (g_P F^p)$ & $ 
        2E^2 \vec p\cdot [ g_V^e  ( \vec s_+ - \vec s_-) - g_A^e (   h_+ \vec s_-  +  h_- s_+) ]$\\
${\rm Re}(g_P F^p)$ & $ 
        2E \vec K \cdot [ g_A^e   ( \vec s_+ - \vec s_-) -   g_V^e 
(h_+ \vec s_-  + h_- \vec s_+) ]\times \vec p$\\
${\rm Im}(g_S F^p)$ & $ 
        2E \vec K \cdot [ - g_V^e   ( \vec s_+ + \vec s_-) -   g_A^e 
(h_+ \vec s_-  - h_- \vec s_+) ]\times \vec p$\\
   ${\rm Re} (g_S F^p)$ & $ 
        2E^2 \vec p\cdot [ g_A^e  ( \vec s_+ + \vec s_-) + g_V^e (   h_+
\vec s_-  -  h_- \vec s_+) ]$\\
       ${\rm Im} (g_P F^s)$ & $2E^2\frac{E_p}{Pm} \vec p \cdot  [ 
g_V^e (\vec s_+ -\vec s_-) - g_A^e (
h_+ \vec s_- +  h_- \vec s_+ ) ]$\\
      ${\rm Re} (g_P F^s)$ &$ 2E \frac{E_p}{Pm} \vec K\cdot  [ g_A^e  (
\vec s_+ - \vec s_-) - g_V^e ( h_+ \vec s_- + h_- \vec s_+)
           ]\times \vec p$\\
       ${\rm Im} (g_S F^s)$ & $2E \frac{E_p}{Pm} \vec K \cdot [ - g_V^e ( \vec
s_+ + \vec s_-)  - g_A^e  (h_+ \vec s_- - h_- \vec s_+) 
           ]\times \vec p $\\
       ${\rm Re} (g_S F^s)$ & $2E^2\frac{E_p}{Pm} \vec p \cdot  [
g_A^e (\vec s_+ +\vec s_-) + g_V^e (
h_+ \vec s_- -  h_- \vec s_+ ) ]$\\
       $ {\rm Im} (g_P F^t) $ & $2E^2\cot\theta \frac{1}{P} \vec p \cdot
[ -g_V^e (\vec s_+ - \vec s_-) + g_A^e ( h_+ \vec s_- + h_- \vec s_+)
          ]$\\
       $ {\rm Re} (g_P F^t)$ &$ 2E \cot\theta  \frac{1}{P} \vec K \cdot
[- g_A^e (\vec s_+ - \vec s_-) + g_V^e (  h_+ \vec s_- +  h_- \vec s_+ )
          ]\times \vec p $\\
      ${\rm Im} (g_S F^t)$& $2E \cot\theta \frac{1}{P} \vec K \cdot  [   
 g_V^e ( \vec s_+ + \vec s_-) + g_A^e (h_+ \vec s_-  -  h_- \vec s_+ )
        ] \times \vec p$\\
       $ {\rm Re} (g_S F^t) $& $2E^2 \cot\theta \frac{1}{P} \vec p \cdot
[ - g_A^e ( \vec s_+ + \vec s_-) 
-  g_V^e ( h_+  \vec s_- - h_- \vec s_+) ]$\\
       $ {\rm Im} (g_P F^n)$&$ 2E \csc\theta \frac{1}{P} \vec K \cdot  [     
 g_V^e ( \vec s_+ - \vec s_-)  - g_A^e ( h_+ \vec s_- + h_- \vec s_+)  
         ]\times \vec p $\\
       $ {\rm Re} (g_P F^n) $ & $2E^2 \csc\theta \frac{1}{P}  \vec
p\cdot [ - g_A^e ( \vec s_+ - \vec s_-) + g_V^e (h_+ \vec s_- + h_- \vec s_+)
           ]$\\
       $ {\rm Im} (g_S F^n) $ & $ 2E^2 \csc\theta \frac{1}{P} \vec p \cdot  [ 
 g_V^e (s_+ + s_-)  + g_A^e ( h_+ \vec s_- - h_- \vec s_+ ) 
 ]$\\
       $ {\rm Re} (g_S F^n) $ & $ 2E \csc\theta \frac{1}{P} \vec K \cdot   [   
 g_A^e (\vec s_+ + \vec s_-) + g_V^e (h_+ \vec s_- - h_- \vec s_+)
        ] \times \vec p$\\
\hline
\end{tabular}
\end{center}
\caption{Correlations due to structure functions for scalar and
pseudo-scalar
BSM physics 
the interference of the SM amplitude with the BSM physics.  An overall
factor of 8 has been suppressed to be consistent with published
results, refs.~\cite{BASDR1,BASDR2}.
Note the symmetry of the correlations under the simultaneous interchange
of Im$\leftrightarrow$Re, $g_V^e \leftrightarrow g_A^e$.
}\label{scalartable}
\end{table}

\begin{table}
\begin{tabular}{|l|c|}
\hline
Structure function & Correlation \\
\hline
${\rm Re} ( W_1)$ &
       $ 4 E^2   [    \vapav (h_+ - h_-) -  \vvpaa (h_+ h_- - 1) ]$\\&\\$
  {\rm Im} ( W_2^{pp})    $ & $
       2 E     \vamav [ \vec p \cdot \vec s_+ \vec p \cdot (\vec K \times
\vec s_-)  +  \vec p \cdot \vec s_- \vec p \cdot (\vec K \times \vec s_+
) ] $\\&\\$
%
%
  {\rm Re} ( W_2^{pp})    $ & $
        E^2  [ 
        2\vec p^T\cdot \vec p^T    \{  \vapav (h_+ - h_-) - \vvpaa
(h_+ h_- - 1) 
$\\&$
+ \vvmaa \vec s_-\cdot \vec s_+  \} 
 - 4 \vvmaa  \vec p \cdot \vec s_+ \vec p \cdot \vec s_- ]
$\\&\\$


  {\rm Im} ( W_2^{ps})    $ & $
        E \frac{E_p}{Pm} 2 \vamav  [  \vec p \cdot \vec s_+ \vec p \cdot
(\vec K \times \vec s_-)  +  \vec p \cdot \vec s_- \vec p \cdot
(\vec K \times \vec s_+ ) ]
$\\&
\\$
  {\rm Re} (W_2^{ps})    $ & $
          E^2 \frac{2E_p}{Pm}  [   \vec p^T\cdot \vec p^T    
\{   \vapav (h_+ - h_-) - \vvpaa (h_+ h_- - 1) 
  $\\&$      
+ \vvmaa  \vec s_-\cdot \vec s_+ \} 
- 2 \vvmaa \vec p \cdot \vec s_+ \vec p \cdot \vec s_- ]
 $\\&\\$

   {\rm Im} ( W_2^{pn})    $ & $
         E^2 \csc\theta \frac{2}{P}   \vamav     [ \vec p^T \cdot \vec
p^T  \vec s_+
\cdot \vec s_- - 2 \vec p \cdot \vec s_+ \vec p \cdot \vec s_-] 
$\\&\\$
  {\rm Re} (W_2^{pn})    $ & $
         E \csc\theta \frac{2}{P}   \vvmaa  [  -  \vec p \cdot \vec s_+ \vec p \cdot (\vec K \times \vec s_-)  
-  \vec p \cdot \vec s_- \vec p \cdot (\vec K \times \vec s_+ )  ]
$\\&\\$
  {\rm Im} ( W_2^{pt})    $ & $
        E \cot\theta \frac{2}{P} \vamav  [  - \vec p \cdot \vec s_+ \vec p \cdot (\vec K \times \vec s_-) 
-  \vec p \cdot \vec s_- \vec p \cdot (\vec K \times \vec s_+ )  ]
$\\&\\$
%
  {\rm Re} (W_2^{pt})    $ & $
         E^2\frac{2}{P}   \cot\theta [ \vec p^T \cdot \vec p^T  \{ 
-\vapav   (h_+ - h_-) 
+ \vvpaa (h_+ h_- - 1) 
$\\&$
 - \vvmaa \vec s_-\cdot \vec s_+  \} 
+    2 \vvmaa \vec p \cdot \vec s_+ \vec p \cdot \vec s_-  ]
$\\&\\

$
  {\rm Im} (W_3^{pq})    $ & $
        8 P E^3 \cos\theta  [   \vvpaa (h_+ - h_-) - \vapav (h_+ h_- - 1)]

$\\
 &\\
$
  {\rm Im} (W_3^{ps})    $ & $
        4 E^2 m \cos\theta  [ - \vvpaa (h_+ - h_-) + \vapav (h_+ h_- - 1)]
$\\
 &\\
$
  {\rm Im} ( W_3^{qs})    $ & $
        8 E^3\frac{E_p}{ m} \cos\theta  [ - \vvpaa (h_+ - h_-) +
\vapav (h_+ h_- - 1)]

$\\
 &\\
$
  {\rm Im} ( W_3^{pt})    $ & $
        4 E^2E_p \sin\theta  [ - \vvpaa (h_+ - h_-) + \vapav (h_+ h_- - 1)]
$\\
 &\\
$
  {\rm Im} ( W_3^{qt})    $ & $
        8 E^3 \sin\theta  [ - \vvpaa (h_+ - h_-) + \vapav (h_+ h_- - 1)]
$\\
 &\\
  ${\rm Im} (W_4^{pn})    $ & $
       2 E^2 P \sin\theta   [ -  (g_Vg_V^e+g_Ag_A^e) (h_+ - h_-) +
(g_Vg_A^e+g_Ag_V^e) (h_+ h_- - 1) ]
$\\
\hline
\end{tabular}
\caption{As in Table \ref{scalartable}, for vector and axial-vector
couplings }\label{vectortable}
\end{table}

\begin{table}
	\begin{center}
\begin{tabular}{|c|c|}
\hline
Structure function & Correlation \\
\hline
Re($g_T$\Tone	 )& $

        8E^2P\cos\theta \vec K \times \vec p \cdot [
   g_V^e \splus -  g_A^e \hsminus ]
$\\

Re($g_T$\Ttwo	 )& $ 4Em\cos\theta \vec K \times \vec p \cdot [- g_V^e \splus
+ g_A^e \hsminus ]$\\
 
Re($g_T$\Tthree       )& $4E^2  
        P  \sin\theta \vec p\cdot [g_V^e \sminus - g_A^e \hsplus ]

   $\\  Re($g_T$\Tfour  )&$
4 EE_p  \sin\theta     \vec K \times \vec p \cdot [- g_V^e \splus 
+ g_A^e \hsminus ]

$\\  
    Re($g_T$\Tfive  )&$
        8E^2\frac{E_p }{ m}\cos\theta \vec K \times \vec p \cdot [
  - g_V^e \splus +  g_A^e \hsminus ]

   $\\  Re($g_T$\Tsix  )&$
0 

   $\\  Re($g_T$\Tseven  )&$
        8E^2 \sin\theta \vec K \times \vec p \cdot [
 - g_V^e \splus +  g_A^e \hsminus ]

   $\\  Re($g_T$\Teight  )&$
        8E^2 \frac{E_p }{ m}\cos\theta \vec K \times \vec p \cdot [
   g_V^e \splus -  g_A^e \hsminus ]

   $\\  Re($g_T$\Tnine  )&$
       8 E^3 \cot\theta \vec   p\cdot [  g_V^e \splus - g_A^e \hsminus ]

   $\\  Re($g_T$\Tonezero  )&$
8 E^2 \cos\theta \cot\theta   \vec K \times \vec p \cdot 
[ - g_V^e \splus +  g_A^e \hsminus ]

   $\\  Re($g_T$\Toneone  )&$ 
        4E \vec K \times \vec p \cdot [ - g_V^e \sminus +  g_A^e \hsplus ]

   $\\  Re($g_T$\Tonetwo  )&$

0

   $\\  Re($g_T$\Tonethree  )&$
        4 E \frac{E_p}{Pm} \vec K \times \vec p \cdot [
  - g_V^e \sminus +  g_A^e \hsplus ]

   $\\  Re($g_T$\Tonefour  )&$
        4 E^2 \csc\theta (\frac{1}{P}) \vec  p\cdot [-
  g_V^e \sminus +  g_A^e \hsplus ]

   $\\  Re($g_T$\Tonefive  )&$
        4 \cot\theta E \frac{1}{P} \vec K \times \vec p \cdot 
[  g_V^e \sminus -  g_A^e \hsplus ]

   $\\  Re($g_T$\Tonesix  )&$
        8 E^3 P \cos\theta \vec   p\cdot [-
   g_V^e \sminus +  g_A^e \hsplus ]

   $\\  Re($g_T$\Toneseven  )&$
        4E^2 m \cos\theta \vec  p\cdot [
   g_V^e \sminus -  g_A^e \hsplus ]

   $\\  Re($g_T$\Tthreesix  )&$

4 P E \sin\theta \vec K \times \vec p \cdot [ - g_A^e \hsminus
+  g_V^e  \splus ]

   $\\  Re($g_T$\Tthreeseven  )&$
4 E_p E^2 \sin\theta \vec p \cdot [- \ga \hsplus
+ \gv \sminus )]

   $\\  Re($g_T$\Toneeight  )&$
        8 \frac{E_p}{m} E^3  \cos\theta \vec   p\cdot [
    g_V^e \sminus -  g_A^e \hsplus ]

   $\\  Re($g_T$\Tthreeeight  )&$
	0

   $\\  Re($g_T$\Tthreenine  )&$
8 E^3 \sin\theta \vec p\cdot  [- \ga \hsplus  
+ \gv \sminus ]

   $\\  Re($g_T$\Tonenine  )&$
        8 \frac{E_p}{m} E^3  \cos\theta \vec   p\cdot [
  - g_V^e \sminus +  g_A^e \hsplus ]

   $\\  Re($g_T$\Ttwozero  )&$
        8 \cot\theta E^2 \vec K \times \vec p \cdot 
[  g_V^e \sminus -  g_A^e \hsplus ]

   $\\  Re($g_T$\Ttwoone  )&$
        8 E^3 \cos\theta\cot\theta \vec p\cdot 
[   g_V^e \sminus -  g_A^e \hsplus ]
         
   $\\  Re($g_T$\Ttwotwo  )&$
        4 E^2  \vec p\cdot 
[   g_V^e \splus -  g_A^e \hsminus ]
         
   $\\  Re($g_T$\Ttwothree  )&$
0

   $\\  Re($g_T$\Ttwofour  )&$
        4 E^2 \frac{E_p}{Pm} \vec p\cdot 
[   g_V^e \splus -  g_A^e \hsminus ]
         
   $\\  Re($g_T$\Ttwofive  )&$
        4 \csc\theta E \frac{1}{P} \vec K \times \vec p \cdot 
[ -  g_V^e \splus +  g_A^e \hsminus ]
         
   $\\  Re($g_T$\Ttwosix  )&$
        4 E^2 \frac{1}{P}\cot\theta \vec p\cdot 
[ - g_V^e \splus +  g_A^e \hsminus ]
         
$\\
\hline
\end{tabular}
	\end{center}
\caption{As in Table \ref{scalartable} for real parts of the tensor couplings}
\label{retensortable}\end{table}

\begin{table}
	\begin{center}
\begin{tabular}{|c|c|}
\hline
Structure function & Correlation \\
\hline
Im($g_T$\Tone )& $
        8 P E^3 \cos\theta \vec p\cdot 
[  g_A^e \splus -  g_V^e \hsminus ]
         
   $\\  Im($g_T$\Ttwo )&$
        4 m E^2 \cos\theta \vec p\cdot 
[  - g_A^e \splus  +   g_V^e \hsminus ]
         
   $\\  Im($g_T$\Tthree )&$
        4 \sin\theta P E \vec K \times \vec p \cdot
[  g_V^e \hsplus -  g_A^e \sminus ]

   $\\  Im($g_T$\Tfour )&$
        4 \sin\theta E_p E^2  \vec p \cdot
[  g_V^e \hsminus -  g_A^e \splus]

   $\\  Im($g_T$\Tfive )&$
        8 \cos\theta \frac{E_p}{m} E^3\vec  p \cdot
[  g_V^e \hsminus -  g_A^e \splus]

   $\\  Im($g_T$\Tsix )&$
0

   $\\  Im($g_T$\Tseven )&$
        8 \sin\theta  E^3\vec  p \cdot
[  g_V^e \hsminus -  g_A^e \splus]

   $\\  Im($g_T$\Teight )&$
        8 \cos\theta \frac{E_p}{m} E^3\vec  p \cdot
[ -g_V^e \hsminus +  g_A^e \splus]

   $\\  Im($g_T$\Tnine )&$

8 E^2 \cot\theta  \vec K \times \vec p\cdot 
[  g_V^e \hsminus -  g_A^e \splus ]

   $\\  Im($g_T$\Tonezero )&$
8 E^3 \cos\theta \cot\theta \vec p\cdot 
[   g_V^e \hsminus -  g_A^e \splus ]

   $\\  Im($g_T$\Toneone )&$
        4  E^2 \vec p \cdot
[  g_V^e \hsplus -  g_A^e \sminus]

   $\\  Im($g_T$\Tonetwo )&$

0

   $\\  Im($g_T$\Tonethree )&$
        4  E^2 \frac{E_p}{Pm}\vec p \cdot
[  g_V^e \hsplus -  g_A^e \sminus]

   $\\  Im($g_T$\Tonefour )&$

4 E \frac{1}{P}\csc\theta  \vec K \times \vec p\cdot 
[ - g_V^e \hsplus +  g_A^e \sminus ]

   $\\  Im($g_T$\Tonefive )&$

        4 \cot\theta  E^2 \frac{1}{P} \vec p \cdot 
[- g_V^e \hsplus +  g_A^e \sminus]

   $\\  Im($g_T$\Tonesix )&$

        8 \cos\theta E^2 P \vec K \times \vec p \cdot 
[  - g_V^e \hsplus +  g_A^e \sminus ]

   $\\  Im($g_T$\Toneseven )&$

        4 \cos\theta Em \vec K \times \vec p \cdot 
[  g_V^e \hsplus -  g_A^e \sminus ]

   $\\  Im($g_T$\Tthreesix )&$
4 P E^2 \sin\theta\vec p \cdot [  \ga \splus
- \gv \hsminus ]

   $\\  Im($g_T$\Tthreeseven )&$

4 E_p E \sin\theta \vec K \times \vec p \cdot [ - \ga \sminus 
+ \gv \hsplus ]

   $\\  Im($g_T$\Toneeight )&$

        8 \cos\theta E^2 \frac{E_p}{m} \vec K \times \vec p \cdot 
[  g_V^e \hsplus -  g_A^e \sminus ]

   $\\  Im($g_T$\Tthreeeight )&$
	0

   $\\  Im($g_T$\Tthreenine )&$

 8 E^2 \sin\theta \vec K \times \vec p \cdot [ - \ga \sminus
+ \gv \hsplus ]

   $\\  Im($g_T$\Tonenine )&$

        8 \cos\theta E^2 \frac{E_p}{m} \vec K \times \vec p \cdot 
[ - g_V^e \hsplus +  g_A^e \sminus ]

   $\\  Im($g_T$\Ttwozero )&$
8 E^3 \cot\theta \vec p\cdot 
[ - g_V^e \hsplus +  g_A^e \sminus ]

   $\\  Im($g_T$\Ttwoone )&$
        8 \cos\theta \cot\theta  E^2 \vec K \times \vec p \cdot 
[  g_V^e \hsplus -  g_A^e \sminus ]

   $\\  Im($g_T$\Ttwotwo )&$
        4  E  \vec K \times \vec p \cdot 
[   g_V^e \hsminus -  g_A^e \splus ]

   $\\  Im($g_T$\Ttwothree )&$

0

   $\\  Im($g_T$\Ttwofour )&$
        4  E \frac{E_p}{Pm} \vec K \times \vec p \cdot 
[   g_V^e \hsminus -  g_A^e \splus ]

   $\\  Im($g_T$\Ttwofive )&$
        4 \csc\theta  E^2 \frac{1}{P}\vec p \cdot 
[  + g_V^e \hsminus -  g_A^e \splus ]

   $\\  Im($g_T$\Ttwosix )&$
        4 \cot\theta  E \frac{1}{P} \vec K \times \vec p \cdot 
[  - g_V^e \hsminus +  g_A^e \splus ]

$\\
\hline
\end{tabular}
	\end{center}
\caption{As in Table \ref{scalartable} for imaginary parts of the tensor couplings }
\label{imtensortable}\end{table}

In the absence of any further assumptions of the theory, it is not
possible to draw very pointed conclusions. However, we can still deduce
some useful points, very often, related to what measurements are not
possible even under the very minimal assumptions made by us.
Examining the tables, one can make the following observations. 
Many of the observations we have here are similar to the
case of one- and two-particle inclusive measurements without
spin resolution~\cite{BASDR1,BASDR2}.  These may be summarized as follows in
bullet form:
\begin{itemize}
\item
In the case of $S$, $P$ and $T$ couplings, 
all the entries in the corresponding tables vanish for unpolarized
beams, or for longitudinally polarized beams. 

\item
Thus at least one beam has to be transversely
polarized to see the interference.

\item
In case of $V$ and $A$ couplings, 
both beams have to be
polarized, or the effect of polarization vanishes.
It is interesting to note that all the correlations in the latter case are
symmetric under the interchange of $\vec s_+$ and $\vec s_-$.

\item In case of $S$, $P$ and $T$  to observe terms
which correspond to combinations like $(h_-\vec s_+ \pm h_+\vec s_-)$, it is
necessary to have at least one beam longitudinally polarized, and the
other transversely polarized. 

\item
It is only the coupling $g_V^e$ which accompanies the
imaginary part of the structure functions in case of $S$, $P$ couplings,
and $g_A^e$ in case of $T$ couplings. Likewise, $g_A^e$ and $g_V^e$
occur with the real parts in these respective cases.

\item
In the case of vector and axial-vector BSM
interactions the structure functions without final-state 
spin measurement which contribute when polarization
is included are the same as the ones which contribute when beams are
unpolarized, provided absorptive parts are neglected. 
We assume here that the final-state particles which are observed are
themselves eigenstates of CP, in which case, 
the imaginary parts of the structure functions contain
absorptive parts of the BSM amplitudes.
In other words, no qualitatively 
new information is contained in the polarized distributions if we
neglect the imaginary parts of the structure functions and do not make a
final-state spin measurement. {\bf However, this situation is changed when
structure functions dependent on final-state spin are included.}
\item
Most BSM interactions are
chirality conserving in the limit of massless electrons, and can
therefore be cast in the form of vector and axial-vector couplings.
Thus, in a large class of contexts and theories, it is possible to
conclude that polarization does not give qualitatively new information,
unless absorptive parts are involved. {\bf Again, the inclusion of spin
measurement of the one-particle inclusive state changes this situation.}
\item
It is possible to conclude that polarization can be used to
get information on absorptive parts of structure functions  of BSM interactions,
which cannot be obtained with only unpolarized beams, and the
final-state spin resolution can be used to obtain information even on
the dispersive parts.\footnote{It is possible to enhance the
sensitivity to BSM interactions with a judicious choice of signs of the
polarization. Thus even when no new structure functions are uncovered by
polarization, the information on structure functions which can be obtained with
polarized beams can be quantitatively better than that obtained with
unpolarized beams. }
\item
In our case, if absorptive parts are included, there is
a contribution from 
${\rm Im}~W_3^{uv}$.
Again, in this case, it
possible to predict the differential cross section for the polarized
case, if the unpolarized cross section is known.
\item
On the other hand, we see that ${\rm
Im}~W_2^{pp}$  contributes only 
for transversely polarized beams. Thus,
to observe these structure functions, it is imperative to have transverse
polarization of both beams.

\item A further point to notice about the
contributions of ${\rm Im}~W_2^{rw}$ is that if $g_V^e = g_V$ and $g_A^e
= g_A$, the contribution vanishes. In other words, if the new physics
contribution corresponds to the exchange of the same gauge boson as the
SM contribution, so that the coupling at the $e^+e^-$ vertex is the
same, even though the final state may be produced through a new vertex,
the contribution to the distribution is zero. Thus, in case of a neutral
final state, where the SM contribution through a virtual photon vanishes
at tree level,  the
observation of ${\rm Im}~W_2^{rw}$ through transverse polarization
could be used to determine the absorptive part of a loop contribution
arising from $\gamma$ exchange. In case of a charged-particle final
state for which both $Z$ and $\gamma$ contribute, such a contribution
would be sensitive to loop effects arising in both these exchanges.
\end{itemize}

{\bf The features mentioned here capture the main reasons for enhancing
BSM physics in the presence of beam polarization, which is the essence
of the studies of refs.~\cite{POWER,BASDR1,BASDR2}.}
\section{CP and T properties of correlations}\label{characterization}

It is important to characterize the C, P and T properties of the various
terms in the correlations, which would in turn depend on the
corresponding properties of the structure functions which occur in them. 

In this context we recall that a similar analysis was done for  the
one-particle inclusive case treated in \cite{BASDR1}. In that case, we
deduced the important result that when the final state consists of
a particle and its anti-particle, it is not possible to have any CP-odd
term in case of $V$ and $A$ BSM interactions. This deduction depended
on the property that in the centre-of-mass frame, the particle and
anti-particle three-momenta are equal and opposite. In the case
of two-particle inclusive distributions, even if the two particles
observed are conjugates of each other, their momenta are not
constrained. Thus it is possible to have CP-odd correlations even in the
$V,A$ case.  In this section, we present an extension of those analyses
to the case at hand, which is the one-particle inclusive case with
spin-resolution.  It may be noted that the work of \cite{BASDR1} is
the simplest possible realization of this framework.  The present work
is a highly non-trivial extension of the work therein, and is based on
the introduction of not just one, but three different spin quantization axes.
It is not possible to anticipate the results of this analysis and thus
the present work is an important extension.  Furthermore, it brings into
the focus the requirement of a dedicated spin analysis of final state particles
at future $e^+e^-$ colliders.

We now come to a more systematic analysis. 
We consider two important
cases, one when the particle $h$ 
in the final state in $e^+e^-\to h X$ is its own conjugate, and
the other when it is not.
We treat these two cases separately.

\subsection{Case A: $h^c = h$}

In this case, 
the particle $h$ is required to be its own conjugate, and is a spin-half
particle, it would be a Majorana fermion, and therefore
uncharged. Then, if $h$ is light (e.g., a Majorana
neutrino), it would escape detection leading to missing energy and
momentum, and the state $X$ would have
to include a pair of charged particles to make possible a measurement of
this state.   On the other hand, if $h$ is heavy (e.g., a heavy Majorana 
neutrino or a neutralino in a supersymmetric theory), it would decay
making it possible to measure it spin with the help of its decay
products.

We first examine the case of scalar and pseudoscalar interactions. 
When the spin of
$h$ is not measured, the distributions in the first two rows of Table 1
being even under CP would be present if the structure function $F^p$
does not violate CP. On the other hand, the distributions in the third
and fourth rows of Table 1, if seen, would measure possible CP violation
in $F^p$.  

If the spin of $h$ is measured, we have to keep in mind that the
spin-dependent structure functions are multiplied by a factor $\lambda$
depending on the spin-quantization axis. When the spin quantization axis
is along
its momentum direction, the dependence of distributions on the CP
properties of $F^s$ is opposite in sign as for $F^p$, since the spin
projection along the momentum 
(which we denote by $\lambda_s$) and the spin itself have opposite P
properties. Since the distributions
are identical in the two cases, except for an additional factor of
$E_p/(Pm)$ in the former case, the distribution with spin measurement
corresponds to CP opposite to that in the case without spin measurement. 

Since under naive time reversal T momentum and spin have the same
behaviour, viz., change of sign, the CPT property will follow the CP
property. We remind the reader that T here denotes naive time reversal,
i.e., reversal of the spin and momenta vectors,
as opposed to genuine time reversal, in which initial and final states
are also interchanged.

The distributions in the case when the spin
quantization axis is $\vec n$ are very similar, 
and the additional factor in this case is ${\rm cosecant} \theta/P$. However, in
this case, the roles of the real and imaginary parts of the structure
functions are reversed, and the distributions have an interchange $g_V^e
\leftrightarrow g_A^e$ relative to the ones for $F^p$ or $F^s$. This has
an important significance because numerically $g_V^e << g_A^e$. Thus,
the distributions which occur with a certain $F^p$ or $F^s$ will have widely
different numerical value as compared to those occurring with the $F^n$
of the same magnitude. Moreover, the vector $\vec n$ as chosen has
exactly opposite C, P and T properties as compared to $\vec s$.
Correspondingly, $\lambda_n$ has also opposite C, P and T properties as
compared to $\lambda_s$.

In case of the spin quantization axis of $h$ being $\vec t$, the
distributions for $F^t$ are related to those of $F^p$ by a factor
$\cot\theta /P$. This changes the CP property of the distribution, since
$\cot\theta$ is odd under CP. In addition, as $\vec t$ has C opposite to
that of $\vec p$, $\lambda_t$ has C=$-1$, whereas its P and T properties
are the same as those of $\lambda_s$, resulting in  
CP$=+1$, and T=$+1$. Thus, the CP properties of the distributions for
$F^t$ remain
opposite to those for the distributions for $F^p$.

To see how a study of the CP properties would be affected by the
experimental configuration, consider the case when the $e^-$ and $e^+$
beams have only transverse polarization, and whose directions are
oppositely directed. This would be a natural scenario in case of
circular colliders, where the $e^-$ and $e^+$ polarizations, because of
synchrotron radiation via the well-known Sokolov-Ternov
effect, are directed perpendicular to the plane of the
trajectories of the particles, and anti-parallel to each other. In this
case, $\vec s_- = - \vec s_+$. Then,  CP-odd distributions arise
in the case of spin-independent structure function $F^p$ for scalar
couplings, and for spin-dependent structure functions $F^s$, $F^n$ and
$F^t$ for pseudoscalar couplings. 

In the case of vector and axial-vector couplings, if the spin of $h$ is not measured,
CP violation can be seen in
distributions associated with Im($g_VW_3^{pq}$) and Im($g_AW_3^{pq}$)
because the factor $\cos\theta$ occurring there is odd under CP.
This CP violation results from absorptive parts of the structure function $W_3^{pq}$, and
is consistent with the what was already remarked in \cite{BASDR1}, that
the observation of CP violation in the case when the spin of
$h$ is not measured requires either an absorptive part to be present, or
the use of transverse beam polarization. We find that this is not true when the spin
of $h$ is observed because  then there are more possibilities of observing
CP violation, again because of the CP-odd factor $\cos\theta$ (or
$\cot\theta$) as in the distributions associated with $W_2^{pt}$,
or because $\lambda_s$ or $\lambda_n$ is odd under CP,
as in the case of $W_2^{ps}$, $W_2^{pn}$ and $W_4^{pn}$. Thus, even in the
absence of absorptive parts, CP violation is observable even without
transverse beam polarization for the structure function $W_4^{pn}$.
Thus, in the
absence of absorptive parts, CP violation is observable even without
transverse beam polarization for the structure function $W_4^{pn}$.
An example of a suggestion for measurement of CP violation in neutralino
production using the spin of the neutralino along the direction $\vec n$
normal to the production plane can be found in \cite{Kittel:2011rk,Choi}.
Another feature seen is a CP-violating contribution associated with 
the structure function $W_2^{pn}$ which survives only in the presence of
transverse beam polarization, but with 
an entirely different kinematic dependence compared to 
any other structure function.

In the tensor case, when spin is not observed, all $F$-type structure
functions are CP even and all $PF$-type ones are CP odd. Of the
spin-dependent structure functions, 
all $F$-type structure functions with one
superscript $s$, $n$ or $t$ are CP odd, whereas all $PF$-type structure
functions with one such superscript are CP even.

Again, if we restrict ourselves to the configuration where $\vec s_+ = -
\vec s_-$, the surviving CP-odd terms of the ones just listed are only those 
of the type  $F_2^{s,n,t}$ and $PF^{pqp}$. 

It is interesting to notice that, with T as usual denoting naive time
reversal, the distributions which correspond to CPT=$+1$ and CPT=$-1$
arise from the two opposite cases where the structure function does not
have an
absorptive part and where it does have an absorptive part. This follows
from the CPT theorem. If the phases
in the definitions of the structure functions are chosen appropriately,
the former would be associated with the real part of the structure
function and the latter with the imaginary part.


\subsection{Case B: $h^c \neq h$}

In the case when $h$ is not self-conjugate, the most interesting case
would be the one where $X \equiv h^c$, i.e., when only $h$ and $h^c$ 
are pair-produced. In that case, ascribing the momentum $p^c$ to $h^c$,
we can write $\vec p = \frac{1}{2} (\vec p - \vec p^c )$ in the c.m. frame, 
so that
under CP, $P$ is invariant, as also $\cos\theta$. 

Looking at Tables 1,  3 and 4, it is clear that in the case of scalar,
pseudoscalar and tensor interactions, when spin of $h$ is not measured,
the only CP-odd correlations are those which 
have a combination $\SM$, which is C odd and P even,  or the combination 
$\HP$, which is C even and P odd. Thus, for the configuration $\vec s_+
= -\vec s_-$, only CP-odd correlations survive.
In the scalar and pseudoscalar case, the 
CP-odd correlation is present for all structure functions with a
pseudoscalar coupling to leptons. 

In the case of vector and axial-vector couplings, there are no CP-odd
correlations. 

If the spin of $h$ is measured, one cannot make definite statements
about CP properties, because the spin of $h^c$ is not measured.

Some more interesting situations are possible in this situation:
When $h$ is not self-conjugate, and when $h$ and $h^c$ are not pair
produced, one can draw conclusions about the CP properties only by
comparing the distributions for an inclusive process with final state $h
+X$ with those for the final state $h^c + X$. We can construct  special
combinations  $\Delta\sigma^\pm$ corresponding
to the sum and difference of the  partial cross sections
$\Delta\sigma$ and $\Delta\bar\sigma$ for production of $h$ and $h^c$
respectively. We could then make statements of the CP properties of
$\Delta\sigma^\pm$ for various structure functions. 
A general discussion in this case is somewhat complicated. However, in
case we restrict ourselves to only tree-level contributions, then, by
unitarity, the effective Lagrangian can be taken to be Hermitian
 and also avoiding complications arising from absorptive parts
generated by loops.
Then the couplings and structure functions contributing to
$\Delta\sigma$ would be complex conjugates of those contributing to
$\Delta\bar\sigma$. Thus, only the real parts of the couplings would
contribute to $\Delta\sigma^+$ and only the imaginary part to
$\Delta\sigma^-$.  It would not be possible to make such a clear-cut
separation if loop effects are included. In this scenario, it would be possible for terms in
distribution which would be CP even in the earlier case of $h=h^c$ to be
CP odd in the combination $\Delta\sigma^-$. This combination would occur
with only imaginary parts of the couplings. On the other hand, terms
which were CP odd in the case $h=h^c$ would be CP odd in the combination
$\Delta\sigma^+$, and real parts of couplings would contribute to it.

\section{Implications for specific processes}\label{exclusive}
In this section we discuss the implications of our framework to specific
processes.  As in the preceding section, since we are specifically
interested in the possibility of CP violating signatures, 
we need to separately consider
the cases of self-conjugate and non-self-conjugate exclusive final states.
\subsection{Case A: $h^c=h$}
The two important cases we consider here are the production of a pair of
heavy neutrinos in a left-right symmetric model \cite{neu1} 
and a pair of neutralinos in a supersymmetric model 
\cite{Kittel:2011rk,Choi} at
electron-positron colliders, both of which can occur in an appropriate
extension of the SM.

\subsubsection{Heavy neutrino production} Since in many theoretical 
scenarios neutrinos are massive Majorana fermions, our formalism will be
applicable to inclusive neutrino production in most theoretical models.
Moreover, many theories incorporate CP violation, which may be relevant
for baryogenesis through leptogenesis.
The neutrino needs to be heavy, so that it decays in the detector so
that its polarization would be measureable. The neutrino may be
accompanied by another neutrino or some other inclusive state.
Since the process does not  occur in
the SM, we will apply our formalism to the interference of two
amplitudes
for the production, one of which would be through the $s$-channel
exchange of the $Z$ boson, non-vanishing only for a $e^-_Le^+_R$ or
$e^-_Re^+_L$ combination of initial states, and the other, may be
through $s$-, $t$-, or $u$-channel exchanges of massive particles. Since
the cases we consider correspond to unpolarized beams, the interfering
second amplitude would also have to have the same initial helicity
combinations, viz., $e^-_Le^+_R$ or $e^-_Re^+_L$, giving essentially
the same results as would a
vector-axial-vector interaction. Consequently, we can
anticipate the resulting distributions from our Table 2. 
If the spin of the final state is not
measured, the relevant rows in Table 2 which incorporate CP violation
correspond to the combinations Im($g_VW_3^{pq}$) and Im($g_AW_3^{pq}$).
In either case, the factor $\cos\theta$ leads to a forward-backward
asymmetry, which is discussed in \cite{neu1}.

It is conceivable that with polarized beams, and/or by studying the
polarization of the neutrinos, more information can be obtained on the
structure of a possible theory of Majorana neutrinos. While our tables
can be used as an indication of what correlations may be useful, the
corresponding structure functions would have to be worked out in the
specific model being tested.

\subsubsection{Neutralino production}
Neutralinos are Majorana fermions and are relevant 
to supersymmetric extensions of the SM.
Again, neutralino production is not a SM process, we again consider
the interference of two amplitudes
for the production, one through the $s$-channel
exchange of the $Z$ boson, 
and the other
through $t$-, or $u$-channel exchanges of massive charged sleptons.
In the absence of beam
polarization, the terms corresponding to $W_4^{pn}$ would indicate CP
violation, which corresponds to neutralino polarization in a plane
normal to the production plane, and is discussed in 
\cite{Kittel:2011rk}. In \cite{Choi} 
a CP-odd forward-backward asymmetry with transverse beam
polarization and without the neutralino spin being measured is
also presented. Our
formalism misses this term because it arises on account of 
$t$- and $u$-channel contributions in the neutralino pair production
process, whereas we consider only $s$-channel processes.

Of related interest is the discussion in ref.\cite{MPF} where
the production of Dirac and  Majorana particles in fermion-antifermion annihilation
is considered in some generality. The main results relate to the symmetry or anti-symmetry 
of the cross section and the polarization of the observed final state when the 
$e^+e^-$ beams are unpolarized or longitudinally polarized.
These correspond to $V, \,A$ type of interactions ($S,\, P$ and $T$ 
interactions requiring transverse beam polarization in order for them to be observable). 
Also, in the case of Dirac fermions, the results require observation of the spins of 
both the fermion and anti-fermion. Hence we cannot make a comparison with the
corresonding results, 
as our formalism concerns only spin measurement of one final-state fermion. 
In the case of Majorana fermions we have found agreement with the symmetry properties
in all cases studied by them, with the exception of the symmetry property of polarization 
in one of the form factors we consider, viz., Re($W_2^{pt}$). 

\subsection{Case B: $h^c\neq h$}
In the case when the produced particle $h$ is not self-conjugate, the
simplest possibility is that $h$ is produced in association with $h^c$,
its conjugate particle. We will restrict ourselves to this possibility,
as a larger inclusive state is complicated to discuss in specific
detail. 

We would like to emphasize that ours is a model-independent approach,
and we would like to elicit general features in our formalism. We do not
include here predictions from individual models for our structure
functions. Nevertheless, we outline an intermediate step in such a
calculation in case of the $hh^c$ final state.
 
Here there are two possibilities, one when the particle $h$ is
a particle in the SM spectrum, and the other when $h$ is new particle in
an extension of the SM. Examples of the former are when $h$ is a charged
massive particle, like the top quark or the tau lepton. As for
extensions of the SM, cases often studied in the literature are
production of excitations of quarks or leptons, 
charginos in supersymmetric theories,
production of heavy quarks or leptons in a model with extra generations
or an extended gauge group, and 
production of Kaluza-Klein partners of SM fermions in extra-dimension theories.
In all these cases, correponding a number of different underlying
theories,  a unified model-independent approach can be used. There are
two possibilities which we consider:

A. Loop-level BSM contribution to $\gamma hh^c$ and $Z h
h^c$ vertices 

In this case, the amplitudes for $hh^c$ production through $s$-channel
$\gamma$ exchange and through $s$-channel $Z$ exchange are
parametrized in a general way in terms of vector, axial-vector and
tensor couplings, with coefficients which are momentum-dependent form factors.
Thus, the structure functions appearing in our formalism, corresponding
to the interference between SM $\gamma$ and $Z$ exchange contributions
with an indirect loop-level BSM contribution, are represented,
still in a relatively model-independent way, in terms of form factors.
The assumption made is that the BSM contribution appears in the loop
contribution to $\gamma hh^c$ and $Z h h^c$ vertices.

This approach can include models mentioned above, for some of which form factors
have been calculated in the past
\cite{Bartl:1997iq,Bartl:1998nn,Illana:1998gp,Ibrahim:2010hv,Arroyo-Urena:2016ygo}.

B. BSM contribution through effective $e^+e^-hh^c$ interactions

This case includes contributions which do not take place through $\gamma
hh^c$ and $Z h h^c$ vertices. Without explicit details of the production
mechanism, the BSM contribution can be represented as general contact
$e^+e^-hh^c$ interactions, which would include all tensor contributions
in a model-independent way. Again, the interference between the SM
contribution and the BSM contact interaction contribution would result
in the structure functions that we use in our analysis. The structure
functions could be calculated in terms of the contact interaction form
factors. Contact interactions in this context have been studied in 
\cite{Schrempp:1987zy,Hagiwara:1989yk,Babich:2000kx}.

The issue of the spin resolution has been studied in the context of $t\bar{t}$ production
in the presence of BSM physics due to an effective Lagrangian characterized exclusively
by its Lorentz signature.  
In ref.~\cite{Ananthanarayan:2010xs,Ananthanarayan:2012ir} 
it was shown that the availability of helicity amplitudes
for both the initial as well as final state particles allows one to obtain 
distributions and to construct suitable asymmetries to probe BSM physics in a manner 
that was not possible when the helicity was unresolved, in contrast
to the work in ref.~\cite{Ananthanarayan:2003wi}.  Furthermore, it was 
also demonstrated that a correspondence could be made between the one-particle 
inclusive distribution and relate the relevant structure functions to the 
parameters of the effective Lagrangian of the exclusive process~\cite{BASDR1}.  
This work was further extended to a scenario of measurements
of the spin in the so-called beam-line and off-diagonal bases 
to enhance the sensitivity to BSM physics~\cite{Ananthanarayan:2012ir},
where these bases were discussed in the context of $t\bar{t}$
production in some detail in ref.~\cite{Parke:1996pr}.

We now study these cases in some more detail.

\subsubsection{A. Loop-level BSM contribution to $\gamma hh^c$ and $Z h
h^c$ vertices}
In this section we study the process 
$e^+ e^- \to f \overline{f}$, where $f$ is a quark or a lepton,
a process which will dominate at the ILC.
We also look at further decays of the final-state fermions when they are
heavy and
the momentum correlations amongst these as  probes of BSM
interactions.  
We concentrate on CP-odd correlations which indicate CP violation and
are therefore important to study. However, these are by no means the only
interesting correlations. CP-even correlations could be used to study
new CP-conserving interactions like magnetic dipole moments.


Early work in this regard was the study by
Couture~\cite{Couture1,Couture2} of $e^+e^-\to \tau^+\tau^-$ in
the presence of dipole moments\footnote{Since $\tau$ leptons particles
are highly unstable, they decay very rapidly, the $\tau$ often into
a $\rho$ or a $\pi$ along with neutrino emission.  In contrast,
if the fermion is a top-quark the
decay is into a $b W$.  Therefore, what one considers in
practice is the possibility of probing the electric dipole moment via momentum
correlations of the decay products, which in reality probes the
spin correlations of the fermion pair $f\bar f$ produced in the reaction, as
the decay is due to the weak interaction which serves as a spin analyzer.}.
Couture has studied in detail the spin-spin correlations in $\tau^+\tau^-$
production and studied the effects of possible electric and magnetic dipole moments on
these, but at energies where the presence of $Z$ boson can
be safely neglected.   It turns out that many important effects
show up only in the presence of the $Z$ boson due to its parity
violating properties, and at energies comparable or significantly
larger than its mass.  Here we consider the process in the full
electro-weak theory, but with a sum over the spin of the $\bar{f}$, 
giving rise therefore to a one-particle inclusive type distribution
with spin resolution, wherein the effective structure functions are
actually known in terms of the dipole strengths.

Consider the process $e^-(p_-)e^+(p_+) \to f(p)\bar f(\bar p)$, in the
presence of anomalous magnetic dipole couplings $\kappa_\gamma$ and $\kappa_Z$
and electric dipole couplings $d_\gamma$ and $d_Z$ of the
spin-half fermion $f$ to $\gamma$ and $Z$, respectively. The effective
Lagrangian of the dipole couplings ($V \equiv \gamma, Z$) is given by
\beq
{\cal L}_{\rm dipole} = \frac{i}{2m_f} \sum_{V=\gamma,Z} 
(\partial_\nu V_\alpha -
\partial_\alpha V_\nu) \bar f
\sigma^{\nu\alpha} (\kappa_V + i d_V \gamma_5) f.
\eeq

The interference term between
the amplitude in the SM model and the amplitude with the anomalous
couplings, summed over the spins of $\bar f$, for the contribution of
one vector boson $V$ ($\gamma$ or $Z$) is
\beq\label{dipoleInt}
\begin{array}{rcl}
M_{\rm SM}  M_{dipole}^* & =&
{\rm Tr}[(1-\g5 \hp + \g5 \sls_+)\slp_+\gamma_\mu(g_V^e-g_A^e \gamma_5)
(1+\g5 h_-+\g5 \sls_-)\slp_-\gamma_\nu]\\
&&    \displaystyle  \times \frac{1}{q^2 - m_V^2}  \frac{1}{m_f}[  g^{\mu\nu}
( q\cdot s  (  - 8 m_f^2 \kappa_V  g_A^f )\\
      && +  4m_f  ( (2 p\cdot q - q^2) i   d_V g_A^f +  q^2  \kappa_V  g_V^f
  ))
       \\&&+ 8p^\mu p^\nu  q\cdot s   (  \kappa_V  g_A^f -  i  d_V g_V^f )
      \\&& + 4(p^\mu s^\nu  + s^\mu p^\nu )    p\cdot q ( - \kappa_V
g_A^f +  i  d_V g_V^f )
\\&&
       -4i(\epsilon^{\alpha\beta\gamma\mu} p_\alpha q_\beta s_\gamma p^\nu + 
       \epsilon^{\alpha\beta\gamma\nu} p_\alpha q_\beta s_\gamma p^\mu)  (  -  k g_V^f +   i d_V g_A^f )
     \\&&  -2i\epsilon^{\alpha\beta\mu\nu}p_\alpha q_\beta [q\cdot s   (  -  \kappa_V  g_V^f +  i
d_V g_A^f ) 
\\&&
         - 4 m_f^2 \kappa_V  g_A^f 
  +  ( 2 p\cdot q - q^2) (\kappa_V  g_V^f - i  d_V g_A^f) 
          ]
      \\&& -2i\epsilon^{\alpha\beta\mu\nu}q_\alpha s_\beta  (  - (2 m_f^2 +p\cdot q) \kappa_V  g_V^f 
+  (p\cdot q -2 m_f^2)i  d_V g_A^f )
      \\&& + 4( p^\mu s^\nu  - s^\mu p^\nu )   (q^2 -  p\cdot q) (\kappa_V
g_A^f -   i  d_V g_V^f  
         )]
\end{array}
\eeq
Here $s$ is the spin four-vector of the fermion $f$, and it would 
represent the helicity four-vector. 
The calculation above is a generic one, and as a result
the vector $s$ can also be replaced
by the vectors $n$ and $t$ to get the results for the cases when the
spin quantization axes are in transverse directions.
Comparing this expression with eqs. (\ref{trace}) and (\ref{Hvector}),
we can immediately identify the various structure functions listed in
eq. (\ref{Hvector}) for the present exclusive case. For this one has
also to consider the $s$ dependent terms above with $s$ replaced by $n$
or $t$. Moreover, some terms in eq. (\ref{dipoleInt}) above do not find
a place in eq. (\ref{Hvector}). To understand this, we note that the
terms with the factors $\epsilon^{\alpha\beta\gamma\mu}p_\alpha q_\beta s_\gamma$
and $\epsilon^{\alpha\beta\gamma\nu}p_\alpha q_\beta s_\gamma$
are vanishing, since in these,
$q$ has only the time component, and the space components of $p$ and $s$
are proportional to each other. When $s$ in these $\epsilon$ tensors
 is replaced by
$n$, $\epsilon^{\alpha\beta\gamma\mu}p_\alpha q_\beta n_\gamma$
reduces, apart from a scalar factor,  
to the transverse spin vector
$t^\mu$, giving a term present in eq. (\ref{Hvector}). 
Similarly,
$\epsilon^{\alpha\beta\gamma\mu}p_\alpha q_\beta t_\gamma$
is proportional to $n^\mu$.

As stated earlier, since the spin of $\bar f$ is not measured, it is not
possible to get correlations which are explicitly even or odd under CP.
However, it is clear that both the CP-even dipole couplings $\kappa_V$
as well as the CP-odd dipole couplings $d_V$ can be measured from  a
measurement of the structure functions through correlations listed in
Table \ref{vectortable}.

In case of the photon, for which $g_A^f$ = 0, the
electric dipole coupling $d_\gamma^f$ appears only in the structure
functions $W_2$, and $W_4$, and then, only in their imaginary parts.  
However, as observed earlier, in case of Im~$W_2$, the coupling
constant combination multiplying the final distribution is $g_V^fg_A^e - g_A^f
g_V^e$, which for the case of photon couplings is 0, as both $g_A^e$ and
$g_A^f$ vanish. Hence the only possibility of measurement of the photon
electric dipole coupling is through $W^4$. In that case, both
longitudinal polarization of the beams and measurement of the transverse
polarization of the final-state fermion is required.

As for the dipole coupling of $f$ to the $Z$, the fact that the vector
coupling $g_V^e$ of the electron to the $Z$ is numerically small
($g_V^e/g_A^e \approx 0.08$) would
play a role in deciding the approximate nature of the distributions. In
case the $f$ is a charged lepton like $\tau$, the corresponding vector
coupling $g_V^f$ to the $Z$ is also small. In that case, neglecting
vector couplings of the $e$ and $\tau$, an examination
of eq. (\ref{dipoleInt}) and Table \ref{vectortable} reveals that the
measurement of the weak dipole moment would require longitudinal
polarization of the $e^-$ or $e^+$ beam, preferably both beam
longitudinally polarized with helicities of opposite sign. 
These observations are in accordance with early work that
pointed out that availability of beam polarization would
significantly enhance the sensitivity to the measurement
of dipole moments~\cite{BASDR3,BASDR4} which generalized
the results for unpolarized beams, see ref.~\cite{BLMN}.

\subsubsection{BSM physics with effective operators in $e^+e^-\to f\bar{f}$}

In \cite{Ananthanarayan:2012ir} we had obtained kinematic distributions
for the process $e^+e^-\to t\bar{t}$ (which would be applicable in
general to any inclusive state $f\bar{f}$ with a heavy fermion $f$) with
transversely polarized beams and when the polarization of the top quark
in the final state is measured. In that work, four-fermion 
contact interactions were used, following the work of Grzadkowski
\cite{Grzadkowski:1995te}. 
It would seem natural that the present
formalism could be applied to that exclusive process as a special case.
It was observed in
\cite{Ananthanarayan:2012ir} that the distributions obtained by explicit
evaluation of the amplitudes were
consistent with those predicted by appropriate structures in our
formalism.  During the course of these investigations
we find, on closer examination,
that with the restrictions of hermiticity on the
couplings of the contact interactions, the structure functions for the
scalar, pseudoscalar and tensor interactions turn out
to be vanishing.
Therefore, it may appear that
there is some kind of an incompatibility between the current framework
and  exclusive processes with scalar or tensor contact interactions.  Furthermore,
although we did find schematic evidence that they can be
mapped on to one another, the two schemes are actually distinct.
The presence of a `current' of the present framework when related to
the adopted contact interaction framework
imposes too stringent a requirement between the various couplings
thereby leading to the vanishing of the structure functions.   

In the case of vector and axial-vector interactions, however, the
formalism is compatible with even contact interactions. It is thus
possible to predict the distributions in this case using our formalism.
This case was not treated in \cite{Ananthanarayan:2012ir}.

One could also compare the results coming from more complicated exclusive states that
could arise in popular extensions such as techni-pion models, for the process
$e^+e^-\to t \bar{t}\pi_t$~\cite{Yue}.  
Other general considerations of discrete symmetry in processes involving 
top- and b-quarks in the SM as well as in the MSSM, without spin resolution 
may be found in refs.~\cite{Ch1,Ch2,Ch3} to cite a few examples.

It may be fruitful to point out that our aim is to establish a model-independent approach. 
As such, it is not intended to take up specific models. Nevertheless, we have taken up some 
explicit categories of final states, both as inclusive states and exclusive states, 
and with that categorization, we have tried to construct general amplitudes characterized 
by form factors. Thus, within our model-independent approach, we have been as explicit as possible
and have tried to illustrate the power of the approach.  The work here is meant to be
a set of consistency checks on specific BSM models, and not meant to be full-fledged
substitute to important and popular extensions of the SM.
\section{Summary and Discussion}\label{conclusions}

We now present a summary of the motivation, approach and
the main results in this work and provide a discussion of these results.

 The motivation for our work comes from the planned high energy and
high performance $e^+e^-$ machines that are now being considered seriously
at very high levels for construction.  While there are several technical
differences between the various machines under discussion in terms of their
physical layout, technology and detector and accelerator aspects between
the ILC, CLIC, FCee and CEPC, the underlying physics that is sought to be
probed is the same.  These machines are expected to be providing a clear
environment and high statistics environment to study SM particles and their
properties at high precision in the light of LHC discoveries.  
A great deal of work has been carried out
to study the feasibility of improving the degree of polarization both for the
electron as well as positron beams.  
It is therefore highly likely that longitudinally
polarized beams will be commissioned at linear colliders.  It is also possible
to create transverse polarization out of longitudinally polarized beams using spin rotators.
Less work has been done for feasibility studies for 
longitudinally polarized beams
at circular collider.  Whereas such studies are desirable 
from the stand point of
accelerator science, the essential features of the physics with polarized beams
remains the same for both linear as well as circular colliders.

In order to build a serious case for high precision SM and BSM studies,
all valid approaches much be studied as diligently as possible.  Our work is
an important model independent effort in this direction, in order to provide
a no-frills approach to the study of fingerprinting BSM physics at these machines.

A model-independent approach to the study of possible BSM physics is to 
represent 
such effects in terms of the most general vertices allowed by Lorentz
invariance
and gauge invariance in the case of various exclusive processes.  
It may be worth emphasizing that BSM models are are characterized 
by alterations to effective vertices
arising in higher orders from the
integrating out of more massive states of the theory,
which would make specific
predictions for the structure functions of the inclusive framework. 
While we have provided some illustrations of such a correspondence, our 
framework has as an advantage model-independence, also as its 
most important distinguishing feature.

An even more
general approach, which is the one pursued here, is one where only one
SM particle is observed, and the effects of all interactions are
represented, assuming Lorentz invariance,  in terms of the
vectors on hand. These are  primarily momenta of the colliding
particles
in $e^+e^-$ collisions, the polarizations of the incoming particles, 
the momentum and the spin of  the observed particle.

 For ease of reading, we now itemize the important points of this work.

\begin{itemize}

\item

In this work, we have started out by considering the most general terms that can
occur in the interference between SM and BSM physics and construct vertices involving
vectors on hand, namely the momenta and
the directions of transverse polarization 
of the incoming particles and the momenta and spin quantization axes of the observed
spin-1/2 particle, consistent with Lorentz covariance. 
Thus, we have computed the spin and momentum correlations, 
expressed as angular distributions,
in $e^+e^-$ collisions
arising from the interference between the virtual $\gamma$ and $Z$
exchange SM amplitudes and BSM amplitudes characterized by their Lorentz
signatures, with the unknown physics encoded into structure functions
for a one-particle inclusive measurement with spin resolution.
Transverse and longitudinal beam polarizations are explicitly included.

\item

The generalization of simple one-particle to two-particle inclusive measurement
was also done some years ago, since the availability of a second
vector gives rise to greater possibilities for structure functions.
The availability of a second vector in the form of the spin quantization
direction in turns gives rise to more intriguing possibilities which
is the subject of this work.  Whereas it is natural to assume the
direction of the motion of the detected particle as the quantization
axis, the full reconstruction of the density matrix of a spin-1/2 particle
requires three mutually orthogonal directions.
We have employed three mutually orthogonal directions as quantization
axes corresponding to longitudinal, normal and transverse directions
in the plane of production.  

\item

A large number of structure functions have to be  introduced, which are taken
to be complex, with definite implications for their properties under 
the discrete symmetries
of C, P and T for the dispersive (real) and absorptive (imaginary) parties, as dictated
by the CPT theorem.   The spin-quantization axes themselves 
can be expressed in terms of the momentum vectors on hand, and allow us to
express the results for the spin-momentum correlations and 
spin-spin correlations in terms of economical tables, namely Tables 1-4.  
While we present the results for the general couplings 
$g_A^e$ and $g_V^e$ directly applicable to $Z$, those
case for the photon are obtained by simply setting $g_V^e=e$ and $g_A^e
= 0$.

\item
Some salient features of the entries in the tables were the following.

\begin{itemize}

\item
Indeed, as in the case of one- and two-particle inclusive study with no
spin resolution, with $S, P$ and $T$ type couplings, transverse polarization
of at least one of the beams is needed to uncover their presence at leading
order, or a hybrid of longitudinal
and transverse polarization.  In the case of imaginary parts of $S$ and $P$ type
structure functions, the result is accompanied by $g_V^e$ while in the case
of $T$ type structure functions it is accompanied by $g_A^e$.  In the
case of the real parts, the vector and axial-vector couplings will have
to be swapped, in accordance with the 
symmetry of the tables under the simultaneous 
swap of vector and axial-vector couplings, and of 
real and imaginary parts of the structure functions.

\item
The structure functions corresponding to $V$ and $A$ type BSM interactions
lead to correlations that have distinctly different properties.  In particular,
without final-state spin resolution beam polarization does not lead to any qualitatively
different  information when the imaginary parts are disregarded.  However,
when spin resolution is included, this is no longer true, which is an
important finding of the present investigation.  In other words,
the appearance of specific structure functions and combinations of initial beam
polarizations which render some spin structure functions as being observable
only with beam polarization is noteworthy.  

\item
Analogously,  
absorptive parts of structure functions with spin resolution are qualitatively different
from those without spin, which was not pointed out earlier.  Thus beam polarization
is crucial for uncovering interactions which cannot be done with unpolarized beams.
Our analysis of the correlations also shows that there are special circumstances
under which contributions may vanish as in the case of a new vector boson, which
would imply equality of the vector- and axial-vector couplings of the electron
and the observed particle, and new interactions would be visible only through loop-effects.

\end{itemize}

\item
As a sequel to the thorough analysis of the general results found in the Tables 1-4,
we have presented a discussion on the nature of the correlations and the 
deductions that can be made on their polarization dependence.
We have also discussed the CP and CPT properties of certain structure
functions.  This was based on a systematic analysis under the rubric
of (a) $h^c=h$, and (b) $h^c\neq h$.  The main features of this analysis
may be summarized as follows:

\begin{itemize}
\item
As one can see from the study without final-state spin being resolved,
when  $h^c=h$ (Majorana fermions), CP violation cannot be observed in
the absence of transverse beam polarization. However, when final spin is
measured, CP violation can  indeed be observed without transverse
polarization of the beams.
Thus, in the
absence of absorptive parts, CP violation is observable even without
transverse beam polarization for the structure function $W_4^{pn}$,
which corresponds to spin measurement of the final-state particle along
the direction of $\vec n$, perpendicular to the production plane.
A realization of this possibility can be found in \cite{Kittel:2011rk}.
It is possible that some other suggestions of different possibilities of
CP-violating distributions discussed here will find realization in other
practical situations.

\item
For the case when the unobserved state $X$ is just $h^c$, that is, $h$
and $h^c$ are pair produced,
in the case of scalar, 
pseudoscalar and tensor interactions, and when spin of $h$ is not measured,
the only CP-odd correlations are those which 
have a combination $\SM$, which is C odd and P even,  or the combination 
$\HP$, which is C even and P odd. Thus, for the configuration $\vec s_+
= -\vec s_-$, only CP-odd correlations survive.
In the scalar and pseudoscalar case, the 
CP-odd correlation is present for all structure functions with a
pseudoscalar coupling to leptons. 
In the case of vector and axial-vector couplings, there are no CP-odd
correlations.

\item
For the case of $h^c\neq h$, interesting results that could
be obtained when $h$ and $h^c$ are pair produced do not directly
apply to the case when only one of them is produced.
It is not possible to make definite statements for the case $h$ and
$h^c$ are produced when the spin of
$h$ is measured, because CP would relate the spin of $h$ to the 
spin of $h^c$ and the latter is not measured.
It is possible to envisage that one has samples separately
with $h X$ and $h^c X$ and construct suitable asymmetries to 
uncover CP violation. 

\end{itemize}

\item
We have provided a  discussion on the possible ways
in which our work can be related to prior studies of exclusive fermion pair production.
In particular, for purposes of illustration, we have considered the process
in the presence of electric and magnetic dipole moments and evaluate the
effective structure functions.  The computation of these
proves to be a useful illustration of the framework.  The
cases of self-conjugate fermions and otherwise have also been
discussed.

\item
We have also 
examined our prior analysis of BSM physics in the form of certain
four-fermion effective operators
with spin resolution in the present framework. We find that in the scalar, pseudoscalar and
tensor cases, our formalism cannot be used without change to the case of
contact interaction framework that had been adopted, which proves
to be too restrictive and leads to vanishing structure functions. 
A sufficiently general inclusive framework that is not based on
`currents' and is more general is yet to be developed and may
prove to be useful for the study where the observed particle is
a boson, in contrast to the present framework where the observed
particle is taken to be a fermion.

\item
The precise modification of the present framework to
other spin bases (as for example  beam-line and off-diagonal bases
considered in the past) is yet to
be analyzed and is work for the future, as also an extension to
spin-spin correlations in a two-particle inclusive state.
\item
In contrast to specific models that go into BSM physics with
exclusive final states, our present work remains very general and model independent,
and could provide the simplest possible framework to study BSM physics to look for
signals independent of assumptions of what lies at higher energy scales.  
Sufficiently precise
data when gathered can be used to study if the structure functions so measured
respect constraints that would be implied by specific extensions of the SM in
exclusive particle production.  

\end{itemize}

Many of the considerations that have been
spelt out for the ILC also apply to the other planned facilities,
namely CLIC, FCee and CEPC.  In particular, our work would also
call for dedicated studies of the advantages of beam polarization
at these facilities, real-life estimates of beam polarization for
transverse as well as longitudinal polarizations that could result
for these configurations, and impact of these on detector design
and accelerator design.

We also suggest that specific studies of such inclusive processes
be implemented also at the level of detector simulations and
event generators to study how departures from ideal detection
can influence the outcome of the concepts put forward here.

\bigskip
\noindent{\bf Acknowledgements:}
BA is partly supported by the MSIL Chair of the
Division of Mathematical and Physical Sciences,
Indian Institute of Science.
SDR acknowledges support from the Department of
Science and Technology, India, under the J.C. Bose National
Fellowship programme, Grant No. SR/SB/JCB-42/2009.

\end{document}